\documentclass[11pt]{article}
\usepackage{graphicx,psfrag}
\usepackage{amsmath,amsthm,amssymb}
\newtheorem{theorem}{Theorem}
\newtheorem{pro}{Proposition}
\newtheorem{lemma}{Lemma}
\newtheorem{remark}{Remark}
\newtheorem{corollary}{Corollary}
\usepackage{color}
\def\hDash{\bot\!\!\!\bot}
\begin{document}
\date{}
\topmargin 0pt
\title
{Model checking for generalized linear models: a dimension-reduction model-adaptive approach
{\small
\author{Xu Guo$^{1,2}$ and Lixing Zhu$^{1}$\footnote{Corresponding author:
Lixing Zhu; email: lzhu@hkbu.edu.hk. The research described here was supported by a grant from the University Grants Council of Hong Kong, Hong Kong. This is a part of Xu Guo's PHD thesis and the finalized version was finished at Nanjing University of Aeronautics and Astronautics. 
}\\
{\small {\small {\it $^1$ Hong Kong
Baptist University, Hong Kong} }}\\
 {\small {\small {\it $^2$ Nanjing University of Aeronautics and Astronautics, Nanjing} }} }} }

\date{}
\maketitle

\renewcommand\baselinestretch{1.5}
{\small

\noindent {\bf Abstract:} Local smoothing  testing  that is based on multivariate nonparametric regression estimation is one of the main model checking methodologies in the literature. However,  relevant tests suffer from the typical curse of dimensionality resulting in slow convergence rates to their limits under the null hypotheses and less deviation from the null under  alternatives. This problem leads tests to not well maintain the significance level and to be less sensitive to  alternatives. In this paper,  a dimension-reduction model-adaptive test is proposed for generalized linear models. The test behaves like a local smoothing test as if the model were univariate, and  can be consistent against any global alternatives and can detect local alternatives distinct from the null at a fast rate that existing local smoothing tests can achieve only when the model is univariate. Simulations are carried
out to examine the performance of our methodology. A real data
analysis is conducted for illustration. The method can readily be extended to global smoothing methodology and other testing problems.
\bigskip

}

\newpage
\section{Introduction}
Consider the following  regression model:
\begin{eqnarray}\label{null}
Y=g(\beta^\tau \mathbf{X}, \theta)+\epsilon,
\end{eqnarray}
here $Y$ is the scalar response, $\mathbf{X}$ is a predictor vector
of $p$ dimension, $g(\cdot)$ is a known squared integrable
continuous function, $\beta$ is any $p$-dimensional unknown
parameter vector, $\theta$ is an unknown parameter of $d$-dimension, and $E(\epsilon|\mathbf{X})=0$.
Generalized linear model is one of its special cases. 

To make statistical inference that is based on regression model
reliable, we should carry out some suitable and efficient model
checking procedures. There are various proposals in the literature for testing model~(\ref{null}) against a general alternative model:
\begin{eqnarray}\label{gen-alter}
Y=G(\mathbf{X})+\epsilon,
\end{eqnarray}
here $G(\cdot)$ is an unknown smooth
function and $E(\epsilon|\mathbf{X})=0.$  Two classes of methods are popularly used: the
local smoothing methods and global smoothing methods. For the former
type, see H\"{a}rdle and Mammen (1993) in which the
$L_2$ distance between the null parametric regression and the
alternative nonparametric regression was considered. Zheng (1996) proposed a
quadratic form conditional moment test which was also independently
developed by Fan and Li (1996). Fan et al. (2001) considered a
generalized likelihood ratio test, and Dette (1999) developed a test that is based on the difference of two variance estimates under the null and alternative respectively. For other developments, see also the
minimum distance test proposed by Koul and Ni (2004) and the
distribution distance test developed by Van Keilegom et al. (2008).
A relevant reference is Zhang and Dette (2004). For global smoothing methodologyies that are based on empirical processes, the examples include the following methods. Stute (1997) introduced nonparametric principal component decomposition that is based on residual marked empirical
process. Inspired by the Khmaladze transformation used in goodness-of-fitting for distributions, Stute et al. (1998b) first developed innovation
martingale approach to obtain some distribution free tests. Stute
and Zhu (2002) provided a relevant reference for generalized linear
models. Khmaladze and Koul (2004) further studied the
goodness-of-fit problem for errors in nonparametric regression. For a comprehensive review, see Gonz\'{a}lez-Manteiga and Crujeiras (2013).

Based on the simulation results in the literature, existing local smoothing methods are sensitive to high-frequency regression models and thus, they often have high power to detect these alternative models. However, a very serious shortcoming is that these methods suffer seriously  from the typical  curse of dimensionality because of inevitable use of multivariate nonparametric function estimation.  Specifically, it results in  that   existing local smoothing-based test statistics under null hypotheses converge to  their limits  at rates  $O(n^{-1/2}h^{-p/4})$ (or $O(n^{-1}h^{-p/2})$ if the test is in a quadratic form)  that are very slow when $p$ is large. The readers can refer to H\"{a}rdle and Mammen (1993) for a typical reference of this methodology. Further, the tests of this type can only detect alternatives distinct from the null hypothesis at the rates of order $O(n^{-1/2}h^{-p/4})$ (see, e.g. Zheng 1996).
This problem has been realized in the literature and there are a number of local smoothing tests that apply re-sampling or Monte Carlo approximation to help determine critical values (or $p$ values). Relevant references include H\"{a}rdle and
Mammen (1993),  Delgado and Gonz\'{a}lez-Manteiga (2001),  H\"{a}rdle et al. (2004), Dette et al. (2007), Neumeyer and Van Keilegom (2010). 
In contrast, though the rate is of order $\sqrt n$, most of existing global smoothing methods depend on  high
dimensional stochastic processes. See  e.g. Stute et al. (1998a).  Because of data sparseness in high-dimensional space, the power performance
often drops significantly.

Therefore, it is of importance to consider how to make local smoothing methods get rid of the curse of dimensionality when  model~(\ref{null}) is the hypothetical model that is actually of a dimension reduction structure. To motivate our method, we very briefly review the basic idea of existing local smoothing approaches. Under the null hypothesis,  $E(Y-g(\beta^\tau \mathbf{X}, \theta)|\mathbf{X})=E(\epsilon|
\mathbf{X})=0$ and under the alternative model~(\ref{gen-alter}), $E(Y-g(\beta^\tau \mathbf{X}, \theta)|\mathbf{X})\not = 0$. Thus, its empirical version with  root-$n$ consistent estimates of $\beta$ and $\theta$ is used as a base to construct test statistics. Its variant is that $E(Y|\mathbf{X})-g(\beta^\tau \mathbf{X}, \theta)=0$. The distance between a nonparametric estimate of $E(Y|\mathbf{X})$ and a parametric estimate of $g(\beta^\tau \mathbf{X}, \theta)$ is a base for test statistic construction (see, e.g. H\"ardle and Mammen 1993). The test that is based on $E(Y-g(\beta^\tau \mathbf{X}, \theta)|\mathbf{X})$ can detect the alternative model~(\ref{gen-alter}).
However, such a very natural idea inevitably involves high-dimensional nonparametric estimation of $E(Y|\mathbf{X})$ or $E(\epsilon|
\mathbf{X})$. This is the main cause of inefficiency in hypothesis testing with the rate of order $O(n^{-1/2}h^{-p/4})$ as aforementioned.

To attack this problem, we note the following fact.
Under the null hypothesis,  it is clear that $E(Y-g(\beta^\tau \mathbf{X}, \theta)|\mathbf{X})=E(\epsilon|
\beta^\tau\mathbf{X})=0$. Thus, it leads to another naive idea to construct test statistic that is based on $E(Y-g(\beta^\tau \mathbf{X}, \theta)|\beta^{\tau}\mathbf{X})$. The test statistic construction sufficiently uses the information provided in the hypothetical model. From the technical development in the present paper, it is easy to see that a relevant test can have the rate of order $O(n^{-1/2}h^{-1/4})$ as if the dimension of $\mathbf{X}$ were $1$. A relevant reference is Stute and Zhu (2002)  when global smoothing method was adopted. But this idea leads to another very obvious shortcoming that  as test statistic is completely based on the hypothetical model, and thus it is actually a directional test rather than an omnibus test. It cannot handle the  general alternative model~(\ref{gen-alter}). 
For instance, when the alternative model is $E(Y|\mathbf{X})=g(\beta^\tau \mathbf{X}, \theta)+g_1(\beta_1^\tau \mathbf{X})$ where $\beta_1$ is orthogonal to $\beta$ and $\mathbf{X}$ follows the standard multivariate normal distribution $N(0, I_p)$. Then $\beta^\tau \mathbf{X}$ is independent of $\beta_1^\tau \mathbf{X}$. When $E(g_1(\beta_1^\tau \mathbf{X}))=0$, it is clear that, still,      under this alternative model $E(Y-g(\beta^\tau \mathbf{X}, \theta)|\beta^\tau\mathbf{X})=0$. Thus, a test statistic that is based on $E(Y-g(\beta^\tau \mathbf{X}, \theta)|\beta^\tau\mathbf{X})$ cannot detect the above alternative. {
Xia (2009) proposed a consistent test statistic by comparing the empirical cross-validation counterparts of the minimum of $E^2(\epsilon-E(\epsilon|\alpha^\tau \mathbf{X}))$ over all unit vectors $\alpha$ with the centered residual sum of squares. When the assumed model is adequate, the centered residual sum of squares should be small. 
However, this procedure cannot provide the corresponding limiting distributions under the null and alternatives and thus cannot test significance at a nominal level. 
As was pointed out by the author himself, under the null hypothesis, the rejection frequency tends to $0$ as $n\rightarrow\infty$. In other words, this method cannot control the significance level in a large sample sense. Further, the computation is also an issue because cross-validation involves intensive computation. The author also provided a single-indexing bootstrap $F$ test. But the consistency of this bootstrap method is not established. Thus, in certain sense it is hard for users to recognize type I and type II error  by this test procedure.}


 Therefore, it is crucial for us to construct a test in the following way:  using sufficiently the information under the null model~(\ref{null}) to avoid dimensionality problem as if the dimension of $\mathbf{X}$ were 1, and adapting to the alternative model~(\ref{gen-alter}) such that the test can detect general alternatives. To achieve this goal, we will apply sufficient dimension reduction technique (SDR, Cook 1998). Define a more general alternative model than model (\ref{gen-alter}). Note that  $\mathbf{X}$ can be rewritten as $B^\tau \mathbf{X}$ where $B=I_p$  an identity matrix. Model~(\ref{gen-alter}) can then be rewritten as $E(Y|X)=\tilde G(B^\tau
\mathbf{X}).$
It is worth noticing that this reformulation  can be for any orthogonal $p\times p$ matrix $B$ because when $G$ is an unknown function, the model can be rewritten as $G(\mathbf{X})=G(BB^{\tau}\mathbf{X}):=\tilde G(B^{\tau}\mathbf{X})$. Thus, $E(\epsilon|\mathbf{X})=0$ is equivalent to  $E(\epsilon|B^\tau \mathbf{X})= 0$, and  $E(G(\mathbf{X})-g(\beta^\tau \mathbf{X}, \theta)| \mathbf{X})\neq 0$ is equivalent to
$E(\tilde G(B^{\tau}\mathbf{X})-g(\beta^\tau \mathbf{X}, \theta)|B^\tau \mathbf{X})\neq 0$. 
In other words, the model can be considered as a
special multi-index model with $p$ indexes. Based on this observation, we  consider a more general alternative model that covers some important models as special cases such as the single-index model and multi-index model:
\begin{eqnarray}\label{gen-alter1}
Y=G(B^{\tau}\mathbf{X})+\epsilon,
\end{eqnarray}
where $B$ is a $p\times q$ matrix with $q$ orthogonal columns for an unknown number $q$ with $1\le q\le p$ and $G$ is an unknown smooth function.  When $q=p$, this model is identical to model~(\ref{gen-alter}), and when $q=1$ and $B=\beta/\|\beta \|$ is a column vector,  model~(\ref{gen-alter1}) reduces to a single-index model with the same index as that in the null model~(\ref{null}). Thus, it offers us a way to construct a test that is automatically  adaptive to the null and alternative models through identifying and consistently estimating $B$ (or $BC$ for an $q\times q$ orthogonal matrix) under both the null and alternatives. From this idea, a local smoothing test will be constructed in Section~2 that has two nice features under the null and alternatives: it sufficiently use the dimension reduction structure under the null and is still an omnibus test to detect general alternatives as existing local smoothing tests try to do. To be precise, the test statistic  under the null converges to its limit at the faster rate of order $O(n^{-1/2}h^{-1/4})$ (or $O(n^{-1}h^{-1/2})$ if the test is in a quadratic form), is consistent against any global alternative and can detect local alternatives distinct from the null at the rate of order $O(n^{-1/2}h^{-1/4})$.   This improvement is significant particularly when $p$ is large because the new test does behave like a local smoothing test as if $\mathbf{X}$ were one-dimensional. Thus, it is expectable that the test can well maintain the significance level and has better power performance than existing local smoothing tests.

The paper is organized by the following way. As sufficient dimension reduction (SDR, Cook 1998) plays a crucial role for identifying and estimating  $BC$ for an $q\times q$ orthogonal matrix $C$, we will give a brief review for it in the next section.
In Section~2,  a dimension-reduction model-adaptive (DRMA) test  is constructed. The asymptotic properties under the null and alternatives are investigated in Section~3.   
In Section~4,  the simulation results are reported and a real data
analysis is carried out for illustration. The basic
idea can be readily  used to other test procedures.
For details, we leave this to Section~5. The
proofs of the theoretical results are postponed to
 the appendix.

\section{Dimension reduction model-adaptive test procedure}

\subsection{Basic test construction}
Recall the hypotheses as: almost surely
\begin{eqnarray}\label{hypo}
H_0: \quad E(Y|\mathbf{X})=g(\beta^{\tau}\mathbf{X}, \theta)\quad \mbox{versus}\quad H_1: \quad E(Y|\mathbf{X})=G(B^{\tau}\mathbf{X}).
\end{eqnarray}
The null and alternative models can be reformulated as: under the null hypothesis, $q=1$ and then  $ B=\tilde
\beta=c\beta$ for some scalar $c$ and under the alternative, $q\geq1$.  Thus,
$$E(\epsilon|\mathbf{X})=0\Longrightarrow E(\epsilon| \beta^\tau \mathbf{X})
=E(\epsilon| B^\tau \mathbf{X})=0.$$  Therefore, under
$H_0$,
\begin{eqnarray}\label{eqn}
&&E(\epsilon E(\epsilon|B^\tau
\mathbf{X})W(B^\tau \mathbf{X}))=E(E^2(\epsilon|B^\tau
\mathbf{X})W(B^\tau\mathbf{X}))=0,
\end{eqnarray}
where $W(X)$ is some positive weight function that is discussed
below.

Under $H_1$,  we have
 \begin{eqnarray}\label{eqn1}
&&E(\epsilon E(\epsilon|B^\tau
\mathbf{X})W(B^\tau\mathbf{X}))=E(E^2(\epsilon|B^\tau
\mathbf{X})W(B^\tau \mathbf{X}))>0,
\end{eqnarray}
 The empirical version of the left hand side in
(\ref{eqn}) can then be used as a test statistic, and $H_0$ can be rejected for large values of the test statistic. 
To this end, we estimate $E(\epsilon|B^\tau \mathbf{X})$ by, when a sample $\{(\mathbf{x}_1, y_1), \cdots, (\mathbf{x}_n, y_n)\}$ is available,
\begin{eqnarray*}
&&\hat E(\epsilon_i|{\hat B}(\hat q)^\tau
\mathbf{x}_i)=\frac {\frac{1}{n-1}\sum_{j\neq i}^n \hat \epsilon_j K_h({\hat B}(\hat q)^\tau \mathbf{x}_i-{\hat B}(\hat q)^\tau \mathbf{x}_j)}{\frac{1}{n-1}\sum_{j\neq i}^n
K_h({\hat B}(\hat q)^\tau \mathbf{x}_i-{\hat B}(\hat q)^\tau \mathbf{x}_j)}.
\end{eqnarray*}
In this formula, $\hat \epsilon_j=y_j-g(\hat\beta ^\tau \mathbf{x}_j,\hat\theta)$,
$\hat\beta$ and $\hat\theta$ are the commonly used least squares estimate of
$\beta$ and $\theta$, 
${\hat B}(\hat q)$ is a
sufficient dimension reduction estimate with an estimated structural dimension $\hat q$ of $q$, $K_h(\cdot)=K(\cdot/h)/h^{\hat q}$ with $K(\cdot)$ being a $\hat q$-dimensional
 kernel function and $h$ being a bandwidth. As the estimations of $B$ and $q$  are crucial to the dimension reduction model-adaption  test (DRMA), we will specify them later.
When the weight $W(\cdot)$ is chosen to be $\hat f(\cdot)$ where
$\hat f({\hat B}(\hat q)^\tau \mathbf{X})$ is a kernel estimate of the
density function $f(\cdot)$ of $B^\tau \mathbf{X}$ and, for any ${\hat B}(\hat q)^\tau \mathbf{x}_i$,
\begin{eqnarray*}
\hat f({\hat B}(\hat q)^\tau \mathbf{x}_i)=\frac{1}{n-1}\sum_{j\neq i}^n
K_h({\hat B}(\hat q)^\tau \mathbf{x}_i-{\hat B}(\hat q)^\tau \mathbf{x}_j).
\end{eqnarray*}
A non-standardized test statistic  is
defined by
\begin{eqnarray}\label{Tn}
V_{n}&=&\frac{1}{n(n-1)}\sum_{i=1}^n\sum_{j\neq i}^n \hat
\epsilon_i\hat \epsilon_jK_h({\hat B}(\hat q)^\tau
(\mathbf{x}_i-\mathbf{x}_j)).
\end{eqnarray}
\begin{remark}
The test statistic suggested by Zheng (1996) is
\begin{eqnarray}\label{Tn_Z}
\tilde{V}_{n}&=&\frac{1}{n(n-1)}\sum_{i=1}^n\sum_{j\neq i}^n \hat
\epsilon_i\hat \epsilon_j\tilde K_h(\mathbf{x}_i-\mathbf{x}_j).
\end{eqnarray}
here $\tilde K_h(\cdot)=K(\cdot/h)/h^p$ with $K(\cdot/h)$ being a $p$-dimensional kernel function. Compared  equation (\ref{Tn}) with equation (\ref{Tn_Z}), there are two main differences. First,  our test  uses ${\hat B}(\hat q)^\tau \mathbf{X}$ in lieu of $\mathbf{X}$ in Zheng (1996)'s test in the classical idea and applies $K_h(\cdot )$ in $V_n$ instead of $\tilde K_h(\cdot)$. This reduces the dimension $p$ down to the dimension $\hat q$.
Second, a further  important ingredient in the test statistic construction is again about the use of kernel function. Under the null, we will show that $\hat q\to 1$, and ${\hat B}(\hat q)\to c\beta$ for a constant $c$, and further  $nh^{1/2}V_n$ has finite limit.
Under the alternative model~(\ref{gen-alter1}), we will show that $\hat q \to q\geq1$ and ${\hat B}(\hat q)\to BC$ for an $q\times q$ orthogonal matrix. The estimation is then adaptive to the alternative model~(\ref{gen-alter1}). This means that the test can be automatically adaptive to the underlying model, either the null or the alternative model.
\end{remark}

\subsection{Identification and estimation of $q$ and $B$}

Note that $B$ is a parameter matrix in the general regression model~(\ref{gen-alter1}). In general, $B$ is not identifiable because for any $q\times q$ orthogonal matrix $C$, $G( B^\tau \mathbf{X})$ can also be written as $\tilde G(C^{\tau} B^\tau \mathbf{X})$. Again, similar to what was discussed before, it is enough to identify $BC$ for an $q\times q$ orthogonal matrix $C$. Thus, what we have to do is to  identify $BC$. To achieve this, we will use the   sufficient dimension
 reduction method(SDR, Cook 1998). Thus, we first briefly review SDR.
Define the intersection of all subspaces $S_B$ spanned by $B$ for all $p\times q$ matrices $B$ such that $Y\hDash
E(Y|\mathbf{X})|B^\tau \mathbf{X}$ where $\hDash$ means ``independent of". Model~(\ref{gen-alter1}) is a special case satisfying this conditional independence. This space, denoted $\emph{S}_{E(Y|\mathbf{X})}$, is called
the central mean subspace (CMS, Cook and Li 2002).
Thus, all we can identify is the subspace $\emph{S}_{E(Y|\mathbf{X})}$ spanned by $B$ rather than $B$ itself in model~(\ref{gen-alter1}). In other words, what we can identify is $\tilde B=BC$ for a $q\times q$ orthogonal matrix or equivalently, $q$ base vectors of $\emph{S}_{E(Y|\mathbf{X})}$. $q$ is called the structural dimension of the central mean subspace $\emph{S}_{E(Y|\mathbf{X})}$. There are several proposals in the literature. The examples include  sliced inverse regression (SIR, Li 1991),
sliced average variance estimation (SAVE, Cook and Weisberg 1991),
contour regression (CR, Li et al. 2005), directional
regression (DR, Li and Wang 2007), likelihood acquired directions
(LAD, Cook and Forzani 2009), discretization-expectation estimation
(DEE, Zhu et al. 2010a) and average partial mean
estimation (APME, Zhu et al. 2010b).  
In
addition,  minimum average variance estimation (MAVE, Xia et al,
2002) can identify and estimate the relevant space with fewer regularity
conditions on $\mathbf{X}$, but requires nonparametric smoothing on involved nonparametric regression function.  As DEE and MAVE have good performance in general, we review these two methods below. 

\subsection{A review on discretization-expectation estimation}
As described in above, our test procedure needs to estimate the
matrix $B$. In this subsection, we first assume the the dimension
$q$ is known in ahead and then we discuss how to select the dimension $q$ consistently.
We first give a brief review of discretization-expectation estimation (DEE), see Zhu et al. (2010a) for details. In sufficient dimension reduction, SIR and SAVE
are two popular methods which involve the partition of the range of $Y$ into several slices and the choice of the number of slices. However, as documented by many authors, for instance, Li (1991), Zhu and Ng (1995) and Li and Zhu (2007), the choice of the number of slices may effect the efficiency and can even yard inconsistent estimates. To avoid the delicate choice of the number of slices, Zhu et al. (2010a) introduced the discretization-expectation estimation (DEE). The basic idea is simple. We first define the new response variable $Z(t)=I(Y\leq t)$, which takes the value 1 if $Y\leq t$ and 0 otherwise. Let $\mathcal{S}_{Z(t)|\mathbf{X}}$ be the central subspace and $\mathcal{M}(t)$ be a $p\times p$ positive semi-definite matrix such that span$\{\mathcal{M}(t)\}=\mathcal{S}_{Z(t)|\mathbf{X}}.$ Define $\mathcal{M}=E\{\mathcal{M}(T)\}$. Under  certain mild conditions, we can have $\mathcal{M}$ is equal to the central subspace that contains the central mean subspace $\mathcal{S}_{E(Y|\mathbf{X})}$. For details, readers can refer to Zhu et al. (2010).

In the discretization step, we construct a new sample $\{\mathbf{x}_i,z_i(y_j)\}$ with $z_i(y_j)=I(y_i\leq y_j)$. For each fixed $y_j$, we estimate $\mathcal{M}(y_j)$ by using SIR or SAVE. Let $\mathcal{M}_n(y_j)$ denote the candidate matrix obtained from a chosen  method such as SIR. In the expectation step, we can estimate $\mathcal{M}$ by $\mathcal{M}_{n,n}=n^{-1}\sum_{j=1}^n \mathcal{M}_n(y_j)$.
The $q$ eigenvectors of $\mathcal{M}_{n,n}$ corresponding to its $q$ largest eigenvalues can be used to form an estimate of $B$. Denote the DEE procedure based on SIR and SAVE be $DEE_{SIR}$ and $DEE_{SAVE}$ respectively. To save space, in this paper, we only focus on these two basic methods. The resulting estimate ${\hat B}(q)$ can be defined by $DEE_{SIR}$ or $DEE_{SAVE}$. Zhu et al. (2010a) proved that ${\hat B}(q)$ is consistent to $BC$ for a $q\times q$ non-singular matrix $C$ and the given $q$.

\subsection{A review on minimum average variance estimation}

As well known, SIR and SAVE needs the linearity and constant conditional variance  conditions for us to successfully estimate $\mathcal{S}_{E(Y|\mathbf{X})}$ (Li, 1991, Cook and Weisberg 1991). In contrast, the minimum average conditional variance
Estimation (MAVE) requires fewer regularity conditions, while asks local smoothing in high-dimensional space. In the following, we also apply MAVE to estimate the base vectors of this subspace, see Xia et al. (2002) for
details. The estimate ${\hat B}$ is the minimizer of
\begin{eqnarray*}
\sum_{j=1}^n\sum_{i=1}^n (y_i-a_j-\mathbf{d}_j^\tau B^\tau
\mathbf{x}_{ij})^2 K_h(B^\tau \mathbf{x}_{ij}),
\end{eqnarray*}
over all $B$ satisfying $B^\tau B=I_{q}$, $a_j$ and $\mathbf{d}_j$,
here $\mathbf{d}_j=G'(B^\tau \mathbf{x}_j)$ and
$\mathbf{x}_{ij}=\mathbf{x}_i-\mathbf{x}_j$. The details of the algorithm can be referred to Xia et al. (2002).
%
The resulting estimate  ${\hat B}(q)$ is also consistent to $BC$ for an $q\times q$ orthogonal matrix when
$q$ is given. When it is unknown, an estimate of $q$ is
involved, which is stated below.

\subsection{Estimation of  the structural dimension $q$}
Consider the DEE-based and MAVE-based estimates. According to  Zhu et al. (2010a),  we determine
$q = \mathrm{dim}(S_{Y|\mathbf{X}})$ by
$$\hat q=\arg\max_{l=1,\cdots,p}
\left\{\frac{n}{2}\times\frac{\sum_{i=1}^l\{\log(\hat\lambda_i+1)-\hat\lambda_i\}}{\sum_{i=1}^p\{\log(\hat\lambda_i+1)-\hat\lambda_i\}}
-2\times D_n\times \frac{l(l+1)}{2p}\right\},$$ where
$\hat\lambda_1\geq \hat\lambda_2\geq\cdots\geq \hat\lambda_p\geq 0$
are the eigenvalues of $\mathcal{M}_{n,n}$, and $D_n$ is a constant to be determined by user.  We should note that the first term in
the bracket can be considered as likelihood ratio, and the second
term is the penalty term with $l(l + 1)/2$ free parameters when the
dimension is $l$. Zhu et al. (2010a) explained the
calculation of this number of free parameters in details. See also
Zhu et al. (2006) for more discussions of this
methodology of BIC type.  The major merit of this methodology is that  the consistency of $\hat q$ only requires
the convergence of $\mathcal{M}_{n,n}$.
Zhu et al. (2010a) proved that under some regularity conditions,
$\hat q$ is a consistent estimate of $q$. Following their suggestion, we choose $D_n=n^{1/2}$.

For  MAVE, we instead suggest a BIC criterion that is a modified version of that proposed by Wang and Yin
(2008), which has the following form:
$$BIC_k=\log(\frac{RSS_k}{n})+\frac{\log(n)k}{\min(nh^k,\sqrt{n})},$$
where $RRS_k$ is the residual sum of squares, and $k$ is the
estimate of the dimension. The form of $RSS_k$ is as follows: $$RSS_k=\sum_{j=1}^n\sum_{i=1}^n
(y_i-\hat a_j-\mathbf{\hat d}_j^\tau {\hat B}(k)^\tau
\mathbf{x}_{ij})^2 K_h({\hat B}(k)^\tau \mathbf{x}_{ij}),$$ here we use
$B(k)$ to denote the matrix $B$ when the dimension is $k$.

The estimated dimension is then
$$\hat q=\min\{l:l=\arg\min_{1\leq k\leq p}\{BIC_k\}\}.$$
Wang and Yin
(2008) showed that under some mild conditions, $\hat q$ is
also a consistent estimate of $q$.

\begin{pro}\label{prop1}
Under the conditions assumed in Zhu et al. (2010a),  the DEE-based estimate $\hat q$ is consistent to $q$. Also under  the conditions assumed in Wang and Yin (2008), the MAVE-based estimate $\hat q$ is consistent to $q$. Therefore, the estimate ${\hat B}(\hat q)$ is a consistent estimate of $BC$ for a $q\times q$ orthogonal matrix $C$.
\end{pro}
It is worth pointing out that these consistencies are under the null and global alternative. Under the local alternatives that will be specified in Section~3, the results are different, we will show the consistency of $\hat q$ to 1.

\section{Asymptotic properties}
\subsection{Limiting null distribution}
Give some notations first. Let $\mathbf{Z}=\beta^\tau
\mathbf{X},\sigma^2(\mathbf{z})=E(\epsilon^2|\mathbf{Z}=\mathbf{z})$, and
 \begin{eqnarray*}
 &&{Var}=2\int K^2(u)du\cdot \int (\sigma^2(\mathbf{z}))^2f^2(\mathbf{z})d\mathbf{z},\\
 &&\widehat{Var}=\frac{2}{n(n-1)}\sum_{i=1}^n\sum_{j\neq i}^n
\frac{1}{h^{\hat q}}K^2(\frac{\hat B(\hat q)^\tau
(\mathbf{x}_i-\mathbf{x}_j)}{h})\hat\epsilon^2_i\hat \epsilon^2_j.
 \end{eqnarray*}
We will show that $\widehat{Var}$ is consistent to $Var$ under the null and local alternatives. This estimation will make the standardized test more sensitive to the alternative.  We now state the asymptotic property of the test statistic under the
null hypothesis.
\begin{theorem}\label{the1}Under $H_0$ and  the conditions in the Appendix, we have
\begin{eqnarray*}
n h^{1/2}V_n \Rightarrow N(0, {Var}),
\end{eqnarray*}
Further, ${Var}$ can be  consistently estimated  by
$\widehat{Var}$.
\end{theorem}

We now standardize $V_n$ to get a scale-invariant statistic.
According to Theorem \ref{the1}, the standardized $V_n$ is

\begin{eqnarray*}
T_n&=&\sqrt{\frac{n-1}{n}}\frac{nh^{1/2}V_n} {\sqrt {\widehat{Var}}}\\
&=&\frac{h^{(1-\hat q)/2}\sum_{i=1}^n\sum_{j\neq i}^n \hat \epsilon_i\hat
\epsilon_jK(\frac{{{\hat B}(\hat q)}^\tau
(\mathbf{x}_i-\mathbf{x}_j)}{h})}{\{2\sum_{i=1}^n\sum_{j\neq i}^n
K^2(\frac{{{\hat B}(\hat q)}^\tau
(\mathbf{x}_i-\mathbf{x}_j)}{h})\hat\epsilon^2_i\hat
\epsilon^2_j\}^{1/2}}.
\end{eqnarray*}
By the consistency of ${\widehat{Var}}$,  the application of the
Slusky theorem yields the following corollary.
\begin{corollary}\label{coro1}
Under the conditions in Theorem 1 and
$H_0$, we  have
$$
T_n^2 \Rightarrow \chi^2_1,
$$
where $\chi^2_1$ is the chi-square distribution with one degree of
freedom.
\end{corollary}
From this corollary, we can then calculate  $p$
values easily  by its limiting null distribution. As a popularly used approach,  Monte Carlo simulation can also be used. We will discuss this in the simulation studies in Section~4.

\subsection{Power study}
We now examine the power performance of the proposed test statistic
under a sequence of  alternatives with the form
\begin{eqnarray}\label{alter}
H_{1n}: Y=g(\beta^\tau \mathbf{X},\theta)+ C_n G(B^\tau \mathbf{X})+\eta,
\end{eqnarray}
where $E(\eta|\mathbf{X})=0$ and the function $G(\cdot)$ satisfies
$E(G^2(B^\tau \mathbf{X}))<\infty$. When $C_n$ is a fixed constant, it is under global alternative, whereas when $C_n\to 0$, the models are local alternatives. In this sequence of models,
$\beta$ is one of the columns in $B$.

Denote $m(\mathbf{X},\beta,\theta)=\textrm{grad}_{\beta,\theta}(g(\beta^\tau
\mathbf{X},\theta))^\tau$, $H(\mathbf{X})=G(B^\tau \mathbf{X})m(\mathbf{X},\beta,\theta)$ and $\Sigma_x=E(m(\mathbf{X},\beta,\theta)m(\mathbf{X},\beta,\theta)^\tau)$.

Before  presenting the main results about the power performance under
the alternative~(\ref{alter}), we give the results about the estimator $\hat q$ of $q$ under the  local alternatives. This is because the local alternative converges to the null model and thus, it is expected that $\hat q$ would also tend to $1$.

\begin{theorem}\label{theo2}
Under  the local
alternatives of (\ref{alter}), when the conditions in  Appendix hold, we have that $\hat q\rightarrow 1.$ Here $\hat q$ is either the DEE-based estimate or the MAVE-based estimate.
\end{theorem}

Now, we are ready to present the power performance.
\begin{theorem} \label{theo3}
Under the conditions in Appendix,  we have the following.  \\
(i) Under the global alternative of \ref{gen-alter1})
$$T_n/(nh^{1/2})\Rightarrow Constant>0.$$
(ii) Under the local alternatives of (\ref{alter}),
$C_n=n^{-1/2}h^{-1/4}$, $nh^{1/2}V_n\Rightarrow N(\mu,{Var})$ and
 $T^2_n\Rightarrow \chi^2_1(\mu^2/{Var}),$
where \begin{eqnarray*}
\mu&=&E\left[\Big(G(B^\tau \mathbf{X})-m(\mathbf{X},\beta,\theta)^\tau\Sigma^{-1}_xE[H(\mathbf{X})]\Big)^2f(\beta^\tau \mathbf{X})\right],
\end{eqnarray*}
here $ \chi^2_1(\mu^2/{Var})$ is
a noncentral chi-squared random variable with one degree
 of freedom and the noncentrality parameter $\mu^2/{Var}$.
\end{theorem}
\begin{remark}
This theorem reveals two important phenomena when compared with Zheng's test: the new test tends to infinity at the rate of order $nh^{1/2}$ whereas Zheng's test goes to infinity at the rate of order $nh^{p/2}$ a much slower rate to infinity; the new test can detects the local alternatives distinct from the null at the rate of order $C_n=n^{-1/2}h^{-1/4},$ whereas the optimal rate for  detectable local alternatives the classical tests can achieve is $C_n=n^{-1/2}h^{-p/4},$ see e.g. H\"ardle and Mammen (1993) and Zheng (1996).
These two facts shows the very significant improvement of the new test achieved.
 \end{remark}

\section{Numerical analysis}

\subsection{Simulations} \label{sec31}
We now carry out simulations to examine the performance of the
proposed test. Because the situation is similar, we then consider
linear models instead of generalized linear models as the
hypothetical models in the following studies. Further to save space, we only consider the SIR-based DEE procedure and MAVE in the following. In this section, we consider 3 simulation studies. In Study~1, the matrix $B$ is $\beta$ under both the null and alternative. Thus, we use this study to examine the performance  for the models with this simple structure and compare the performance when DEE and MAVE are used to determine $BC$ for an $q\times q$ orthogonal matrix. Study~2 is to check, through a comparison with Stute and Zhu's test (2002), how our test is an omnibus test that has the advantage to detect general alternative models rather than a directional test that cannot do this job. Study~3 is multi-purpose: to check the impact from the dimensionality for both our test and a local smoothing test (Zheng 1996). It is note that we choose Zheng's test for comparison because 1). it has tractable limiting null distribution and we can see its performance when limiting null distribution is used to determine the critical values (or $p$ values); 2). Like  other local smoothing tests, we can also use its re-sampling version to determine the critical values (or $p$ values) to examine its performance. Thus, we use it as a representative of local smoothing tests to make comparison. We also conduct a comparison with  H\"ardle and Mammen's (1993) test in a small scale simulation, the conclusions are very similar and thus the results are not reported in the present paper.

\vspace{3mm}
\emph{Study 1}.  Consider
\begin{eqnarray*}\label{simumod1}
&&H_{11}:Y=\beta^\tau \mathbf{X}+a\cos(0.6\pi \beta^\tau \mathbf{X})+\epsilon;\\
&&H_{12}:Y=\beta^\tau  \mathbf{X}+a\exp\{-(\beta^\tau  \mathbf{X})^2\}+\epsilon;\\
&&H_{13}:Y=\beta^\tau  \mathbf{X}+a(\beta^\tau
\mathbf{X})^2+\epsilon.
\end{eqnarray*}
The value $a=0$ corresponds to the null hypothesis and $a \neq 0$ to
the alternatives. In other words, even under the alternative, the model is still single-indexed.
In the simulation, $\beta=(1,1,\cdots,1)^\tau/\sqrt{p}$,
$\mathbf{X}=(X_1,X_2,\cdots,X_p)^{\tau}$ and $p$ is set to be 8.
The observations $\mathbf{x}_i, i=1, 2, \cdots, n$,  are i.i.d.
respectively from multivariate normal distribution
$N(0,\Sigma_j),j=1,2,$ with
\begin{eqnarray*}\Sigma_1=I_{p\times p}; \quad
\Sigma_2=(0.5^{|j-l|})_{p\times p}.
\end{eqnarray*}
Further, $\epsilon$ follows the standard normal distribution $N(0,1)$. In this simulation study,
the replication time is $2, 000$.

\begin{figure}[htbp]
\centering
\includegraphics[width=11cm,height=10cm]{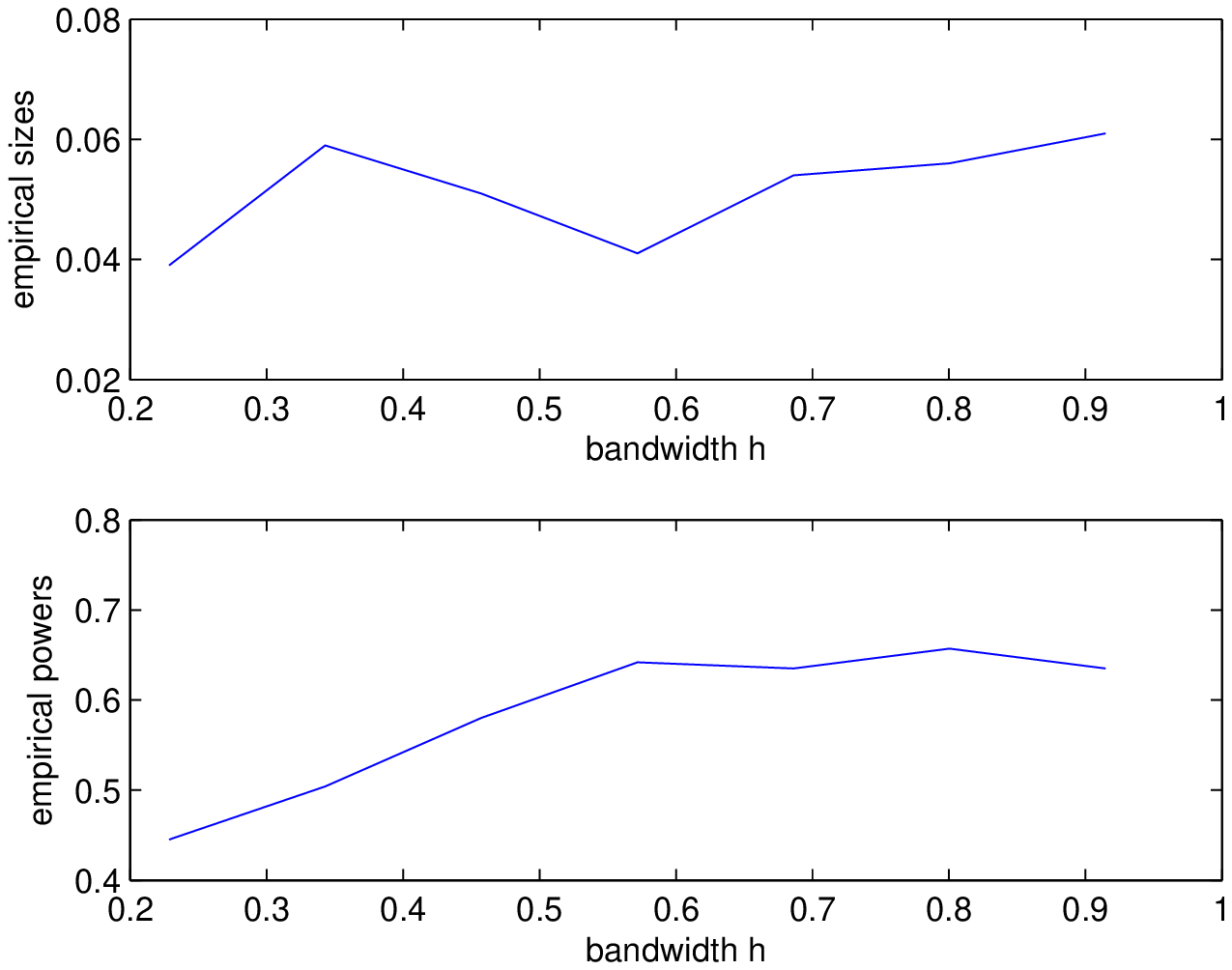}
 \caption{The empirical size and power curves of $T^{DEE}_n$ against the bandwidth $h$ with $X\sim N(0,\Sigma_1)$, $\epsilon\sim
N(0,1)$ and sample size 50 under different choices
of $a$ for study 1 with $a=0$(the above panel) and $a=1$(the below panel). }
 \label{fig1}
\end{figure}

In the nonparametric  regression estimation, the kernel function is taken to be $K(u)=15/16(1-u^2)^2$,
if $|u|\leq 1$; and $0$ otherwise.  The bandwidth is taken to be $h=1.5n^{-1/(4+\hat q)}$ with separately standardized predictors for simplicity. We have this recommended value because to investigate the impact of bandwidth selection, we tried different bandwidth $h$ to be $n^{-1/(4+\hat q)}(0.25+i/4)$ for $i=0,\cdots,8$. For the three models, we found similar pattern we depict in Figure~\ref{fig1} that is under $H_{11}$  with $X\sim N(0,\Sigma_1)$, and the sample size 50 at  the nominal level  $\alpha = 0.05$.   From this figure, we can see that the test is somehow robust to the bandwidth selection when empirical size is a concern: with different bandwidths, our proposed test can control the size  well. On the other hand, the bandwidth selection does have impact for  power performance especially when the bandwidth is too small. Overall, from our empirical experience here, $h=1.5n^{-1/(4+\hat q)}$  is recommendable.

\begin{center}
Figure~~\ref{fig1} about here
\end{center}


Now we turn to study the empirical sizes and powers of our proposed test against alternatives $H_{1i},i=1,2,3$ with the nominal size $\alpha=0.05$. The results are shown in Tables~\ref{tab1}--\ref{tab3}. Write respectively $T^{DEE}_n$ and $T^{MAVE}_n$ as  the corresponding test statistic $T_n$ that is respectively based on DEE and MAVE.  From these three tables, we can have the following observations. In all the  cases we consider, $T^{DEE}_n$ controls the size very well even under the small sample size $n=50$. This suggests that we can rely on  the limiting null distribution to determine critical values (or $p$ values).

For $T^{MAVE}_n$, our simulations which are not reported here show that the empirical sizes intend to be slightly larger than $0.05$. Thus an adjustment on the test statistic is needed. The following size-adjustment would be recommendable:
$$\widetilde{T}^{MAVE}_n=\frac{T^{MAVE}_n}{1+4n^{-4/5}}.$$ The test statistic is asymptotically equivalent to $T^{MAVE}_n$, but has better performance in small or moderate sample size.  The results shown in Tables~\ref{tab1}--\ref{tab3} are with this adjusted test statistic.

As was pointed out in the literature and commented in the previous sections, local smoothing tests usually suffer from dimensionality problem such that they cannot work well in the significance level maintenance with good  power performance. Thus, re-sampling techniques are often employed in finite sample paradigms. A typical technique is the wild bootstrap first suggested by Wu (1986) and well developed later (see, e.g. H\"ardle and Mamenn 1993).
Consider the bootstrap observations: $$y^*_i=\hat\beta^\tau\mathbf{x}_i+\hat\epsilon_i\times V_i.$$
Here $\{V_i\}_{i=1}^n$ is a sequence of i.i.d. random variables with zero mean, unit variance and independent of the sequence $\{y_i,\mathbf{x}_i\}_{i=1}^n$. Usually, $\{V_i\}_{i=1}^n$ can be chosen to be i.i.d. Bernoulli variates with
$$P(V_i=\frac{1-\sqrt{5}}{2})=\frac{1+\sqrt{5}}{2\sqrt{5}},\qquad\,\, P(V_i=\frac{1+\sqrt{5}}{2})=1-\frac{1+\sqrt{5}}{2\sqrt{5}}.$$
Let $T^{*}_n$ be defined similarly as $T_n$, basing on the bootstrap sample $(\mathbf{x}_1,y^*_1)$, $\cdots,(\mathbf{x}_n,y^*_n)$. The null
hypothesis is rejected if $T_n$ is bigger than the corresponding quantile of the bootstrap distribution of $T^{*}_n$. The bootstrap versions of $T_n$ based on DEE and MAVE  are respectively written as $T^{DEE*}_n$ and $T^{MAVE*}_n$.

From Tables~\ref{tab1}--\ref{tab3}, we can see the following. The empirical sizes of $T^{MAVE*}_n$ are slightly larger than $0.05$ especially when $\mathbf{X}\sim N(0,\Sigma_1)$. $\widetilde{T}^{MAVE}_n$ can control the size acceptably in different situations.  $T^{DEE*}_n$ can also maintain the type I error very well, but cannot work better than $T^{DEE}_n$. In summary, $T^{DEE}_n$, $T^{DEE*}_n$, $T^{MAVE*}_n$ and $\widetilde{T}^{MAVE}_n$ all work well. Thus, re-sampling technique seems not necessary for our DRMA tests though $T^{MAVE}_n$ needs some adjustment to get $\widetilde{T}^{MAVE}_n$.
For power performance, Tables~\ref{tab1}--\ref{tab3} suggest that when $\mathbf{X}\sim N(0,\Sigma_1)$, $\widetilde{T}^{MAVE}_n$ generally has higher power than $T^{DEE}_n$ has. Yet, when $\mathbf{X}$ follows $N(0,\Sigma_2)$, $T^{DEE}_n$ becomes the winner. Further,  $T^{MAVE*}_n$ has slightly higher power than $T^{DEE*}_n$  has.  Moreover, for the alternatives $H_{11}$ and $H_{12}$, $T^{MAVE*}_n$ generally has relatively higher power than $\widetilde{T}^{MAVE}_n$ has while for the alternative $H_{13}$, $\widetilde{T}^{MAVE}_n$ is more powerful than $T^{MAVE*}_n$. As for the DEE-based tests, in almost all cases, $T^{DEE}_n$ can have higher power than its bootstrapped version $T^{DEE*}_n$. These  tests are all very sensitive to the alternatives. To be precise, when $a$ increases, the powers can increase very quickly. It seems that the MAVE-based tests tend to be more conservative than the DEE-based tests, and $T^{DEE}_n$ can not only control the size satisfactorily but also have high power in the limited simulations.

\begin{center}
Tables~\ref{tab1}--\ref{tab3} about here
\end{center}
\vspace{3mm}

Further, as commented in Section~1, the test  proposed by Stute and Zhu (2002) is also a dimension reduction test that is however a directional test. Specifically, they developed an innovation process transformation
of the empirical process $n^{-1/2}\sum_{i=1}^n\Big(y_i-g(\hat\beta^\tau \mathbf{x}_i,\hat\theta)\Big) I(\hat\beta^\tau \mathbf{x}_i\leq u)$. By introducing the transformation, the test is  asymptotically distribution-free, but not an  omnibus test though it has been proved to be powerful in many scenarios (see, e.g.  Stute and Zhu 2002; Mora and Moro-Egido 2008). We use the following simulation to demonstrate this claim in the finite sample cases.
\vspace{3mm}

\begin{figure}[htbp]
\centering
\includegraphics[width=11cm,height=10cm]{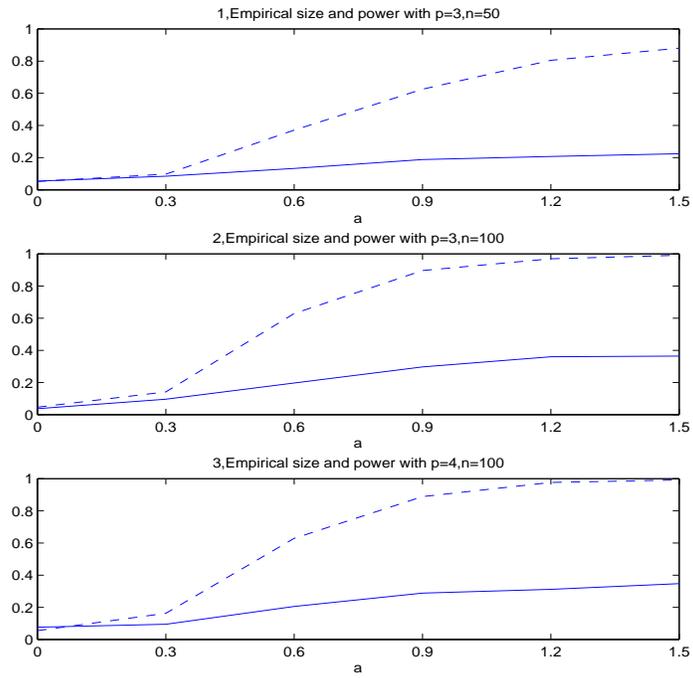}
 \caption{The empirical size and power curves of $T^{SZ}_n$ and $T^{DEE}_n$ in Study~2. The solid and dash line represent the results from $T^{SZ}_n$ and $T^{DEE}_n$ respectively. }
 \label{fig2}
\end{figure}

\emph{Study 2}. The data are generated from the following model:
\begin{eqnarray}\label{modelsim}
Y=\beta^\tau_1 \mathbf{X}+a(\beta^\tau_2
\mathbf{X})^3+\epsilon.
\end{eqnarray}
Consider two cases of $\beta_i$. The first is $\beta_1=(1,0,0)^\tau$, $\beta_2=(0,1,0)^\tau$ for $p=3$; the second one is  $\beta_1=(1,1,0,0)^\tau/\sqrt{2}$, $\beta_2=(0,0,1,1)^\tau/\sqrt{2}$ for $p=4$.
When $p=3$, the sample size $n=50,100$, and when $p=4$, $n=100$. In both the cases,  $X$ and $\epsilon$ are generated from multivariate and univariate  standard normal distribution. Further, consider $a=0.0,0.3,\cdots,1.5$  to examine power performance. Write Stute and Zhu (2002)'s test as $T^{SZ}_n$. To save space, we only present the results of $T^{DEE}_n$ in Figure~\ref{fig2}.  It is obvious that $T^{DEE}_n$ uniformly performs much better than $T^{SZ}_n$.  $T^{SZ}_n$ can have very low powers. In contrast, $T^{DEE}_n$ can efficiently detect the alternatives. Here, we only use this simulation to show $T^{SZ}_n$ is a directional test rather than to show $T^{SZ}_n$ is a bad test. Actually, it has been proved to be a good test in other scenarios (see Stute and Zhu 2002). Further,  for other comparisons, it will be a comparison between local smoothing tests and global smoothing tests. We will have further investigation in an ongoing research.

\begin{center}
Figure~\ref{fig2} about here
\end{center}

To evidence the performance of our test when there are more than one direction under the alternative hypothesis and the impact from dimensionality, we construct the following simulation.

\vspace{3mm}
\emph{Study 3}. The data are generated from the following model:
\begin{eqnarray}\label{modelsim}
Y=\beta^\tau_1 \mathbf{X}+a(\beta^\tau_2
\mathbf{X})^2+\epsilon,
\end{eqnarray}
where $\beta_1=(\underbrace{1,\cdots,1}_{p/2},0,\cdots,0)^\tau/\sqrt{p/2}$, $\beta_2=(0,\cdots,0,\underbrace{1,\cdots,1}_{p/2})^\tau/\sqrt{p/2}$.
Thus, under the null, we have $B=\beta_1$ and under the alternatives,  $B=(\beta_1, \beta_2).$ In this study, we consider $p=2$ and $8$ to examine how the dimension affects Zheng's test and ours.

The observations $\mathbf{x}_i, i=1, 2, \cdots, n$  are generated from multivariate normal distribution
$N(0,\Sigma_j),j=1,2$ and $\epsilon$ from $N(0,1)$
and the double exponential  distribution $DE(0,\sqrt{2}/2)$ with density $f(x)=\sqrt{2}/2\exp\{-\sqrt{2}|x|\}$  with mean zero and variance 1 respectively. To save space, we only consider $T^{DEE}_n$ and $T^{DEE*}_n$ in the following due to their well performance on size control and easy computation.

 When $p=2$, $\beta_1=(1,0)^\tau$ and $\beta_2=(0,1)^\tau$.  The results are reported in Table~4. From this table, firstly, we can observe that Zheng (1996)'s test $T^{ZH}_n$ can maintain the significance level reasonably in some cases, but usually, lower than it. When the bootstrap is used, $T^{ZH*}_n$ performs better in this aspect. In contrast, both  $T^{DEE}_n$ and $T^{DEE*}_n$ can maintain the significance level very well. In other words, the DRMA test we developed does not need the assistance from the bootstrap approach, while Zheng's test is eager for. For the empirical powers, we can find that generally,  $T^{ZH}_n$ and $T^{DEE}_n$ have higher powers than their bootstrap version. However,  for our tests, the differences are negligible. This again shows that the bootstrap helps little for our test. Whereas Zheng (1996)'s test also needs the help from the bootstrap for power performance. Further,  our tests, both  $T^{DEE}_n$ and $T^{DEE*}_n$ are uniformly and significantly more  powerful than Zheng (1996)'s test and the bootstrap version.

\begin{center}
Table~4 about here
\end{center}

 We now consider the $p=8$ case. The results are reported in Table~\ref{tab5}. It is clear that the dimension very significantly deteriorates the performance of Zheng's test. Table~\ref{tab5} indicates that the empirical size of $T^{ZH}_n$ is far away from the significance level. The bootstrap can help on increasing the empirical size, but still not very close to the level. In contrast,  $T^{DEE}_n$ can again maintain the significance level very well and the bootstrap version $T^{DEE*}_n$ does not show its advantage over $T^{DEE}_n$. Further, the power performance of $T^{ZH}_n$ becomes much worse than that under the $p=2$ case shown in Table~4, even the bootstrap version does not enhance the power, which is much lower than that of  $T^{DEE}_n$. This further shows that $T^{ZH}_n$ is affected by the dimension badly, while $T^{DEE}_n$ is not. The empirical powers of $T^{DEE}_n$ and $T^{DEE*}_n$ with $p=8$ are even higher than those with $p=2$. This dimensionality blessing phenomenon is of interest and worthy of a further exploration in another research.

\begin{center}
Table~\ref{tab5} about here
\end{center}



These findings coincide with the theoretical results derived before: existing local smoothing tests have much slower convergence rate ( of order $n^{-1/2}h^{-p/4}$ or $n^{-1}h^{-p/2}$ if the test is a quadratic form) to their limit under the null and less sensitive to local alternative (at the rate of order $n^{-1/2}h^{-p/4}$ to the null) than the DRMA test we developed. 
The simulations we conducted above shows that the DRMA test can simply use its limiting null distribution to determine critical values  without heavy computational burden, and has high power.

\vspace{3mm}

\subsection{Real data analysis}
\begin{figure}[htbp]
\centering
\includegraphics[width=11cm,height=10cm]{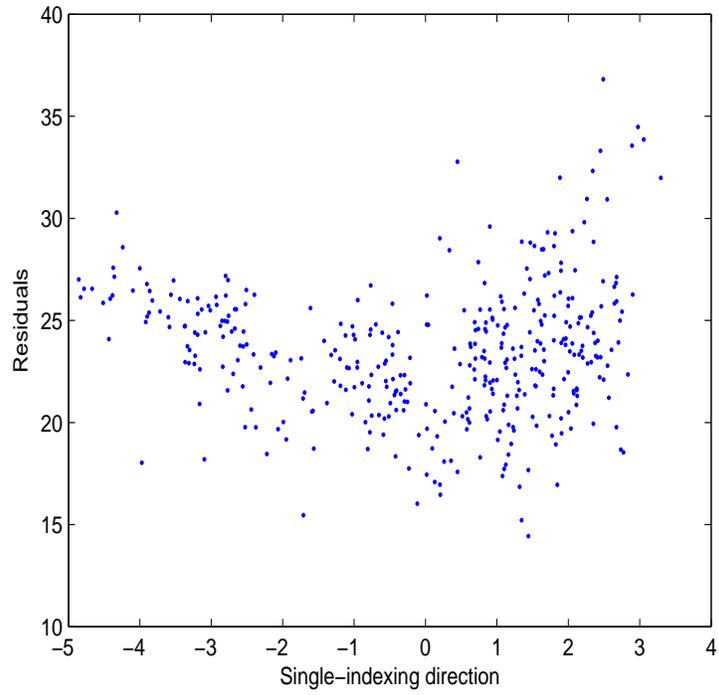}
 \caption{Plot of the residuals from the linear regression model against the single-indexing direction obtained from DEE in the real data analysis.}
 \label{fig3}
\end{figure}

This dataset is obtained from the Machine Learning Repository at the
University of California-Irvine (http://archive.ics.uci.edu/ml/datasets/Auto+MPG).
Recently, Xia (2007) analysed this data set by their method.
The first analysis of this data set is due to Quinlan (1993).
There are 406 observations in the original data set. To illustrate our method, we first
clear the units with missing response and/or predictor and get 392 sample points.
The response variable $Y$ is miles per gallon ($Y$). There are other seven predictors: the number of cylinders ($X_1$), engine
displacement ($X_2$), horsepower ($X_3$), vehicle weight ($X_4$), time to accelerate from
0 to 60 mph ($X_5$), model year ($X_6$) and origin of the car (1 = American, 2 =
European, 3 = Japanese). Since the origin of the car contains more than two
categories, we follow Xia (2007)'s suggestions and define two indictor variables. To be precise, let $X_7 = 1$ if a car
is from America and 0 otherwise and $X_8 = 1$ if it is from Europe and 0 otherwise.
For ease of explanation, all the predictors are
standardized separately. Quinlan (1993) aimed to predict the response in terms of the eight predictors $\mathbf{X}=(X_1,\cdots,X_8)^\tau$. To achieve this goal, a simple linear regression model was adopted. However, as was argued in Introduction, we need to check its adequacy to avoid model misspecification. The value of $T^{DEE}_n$ is $86.5703$ and the p value is $0$. The value of $\widetilde{T}^{MAVE}_n$ is $98.2602$ and the p value is also 0. Hence  linear regression model is not plausible  to predict the response. Figure~\ref{fig3} suggests that a nonlinear model should be used. Moreover, the $\hat q$ is estimated to be 1 by the criterion with DEE in the paper. Thus a single-index model may be applied.
\begin{center}
Figure~\ref{fig3} about here
\end{center}

\section{Discussions}
In this paper, we propose a dimension-reduction model-adaptive test procedure and use Zheng's test as an example to construct test statistic. It is readily extended to other local
smoothing methods discussed in Section~1.
The same principle can be applied  to global smoothing methods. To be
precise, as discussed in Section 2, under the null hypothesis
$Y=g(\beta^\tau \mathbf{X},\theta)+\epsilon$ with $E(\epsilon|\mathbf{X})=0$, we can have
$E(\epsilon|\beta^\tau \mathbf{X})=E(\epsilon|B^\tau \mathbf{X})=0$. While under the
alternative $H_1$, $E(\epsilon|B^\tau \mathbf{X})=E(Y-g(\beta^\tau \mathbf{X},\theta)|B^\tau
\mathbf{X})=G(B^\tau \mathbf{X})-g(\beta^\tau \mathbf{X},\theta)\neq 0$. This motivates us to define
the following test statistics:
\begin{eqnarray*}
R_n(\mathbf{z})={n}^{-1/2} \sum_{i=1}^n\Big(y_i-g(\hat\beta^\tau \mathbf{x}_i,\hat\theta)\Big)
I({{\hat B}(\hat q)}^\tau \mathbf{x}_i\leq \mathbf{z}).
\end{eqnarray*}
This is different from that in Stute and Zhu
(2002) in which ${{\hat B}(\hat q)}=\hat\beta$ in $R_n(\mathbf{z})$ for generalized
linear models. As we commented and compared in theoretical development and simulations, Stute and Zhu's (2002) test is a directional test and then is inconsistent under general alternatives. An ongoing project is doing for this.

Further, extensions of our methodology to missing, censored data and
dependent data set can also be considered. Take the missing response
as an example.  Let $\delta_i$ be the
missing indicator, that is, $\delta_i=1$ if $y_i$ is observed,
otherwise it's equal to zero. Assume that the response is
missing at random. This means
$P(\delta=1|\mathbf{X},Y)=P(\delta=1|\mathbf{X}):=\pi(\mathbf{X})$. For more details, see
Little and Rubin(1987). Again, we consider to test whether the
following regression model holds or not. Namely, $H_0:
Y=g(\beta^\tau \mathbf{X},\theta)+\epsilon$ with $E(\epsilon|\mathbf{X})=0$ and $Y$ is
missing at random. Note that under the null hypothesis
$E(\delta\epsilon/\pi(\mathbf{X})|\beta^\tau
\mathbf{X})=E(\delta\epsilon/\pi(\mathbf{X})|B^\tau \mathbf{X})=0$, while under the alternative
$E(\delta\epsilon/\pi(\mathbf{X})|B^\tau \mathbf{X})\neq 0$. Similarly, we can
construct a consistent test statistic with the following form:
\begin{eqnarray*}\label{Tn_miss1}
V_{n1}&=&\frac{1}{n(n-1)}\sum_{i=1}^n\sum_{j\neq i}^n
\frac{\delta_i}{\hat\pi(\mathbf{x}_i)}\frac{\delta_j}{\hat\pi(\mathbf{x}_j)}\hat
\epsilon_i\hat \epsilon_jK_h({{\hat B}(\hat q)}^\tau
(\mathbf{x}_i-\mathbf{x}_j)),
\end{eqnarray*} here $\hat\pi(\mathbf{x}_i)$ is an estimate, say,
the nonparametric or parametric estimate, of $\pi(\mathbf{x}_i)$,
$\hat\epsilon_i=y_i-g(\hat\beta^\tau \mathbf{x}_i,\hat\theta)$ and $\hat\beta$ and ${{\hat B}(\hat q)}$ is obtained from the complete observed units. Another possible
test statistic takes the following form
\begin{eqnarray*}\label{Tn_miss2}
V_{n2}&=&\frac{1}{n(n-1)}\sum_{i=1}^n\sum_{j\neq i}^n
\delta_i\delta_j\hat \epsilon_i\hat \epsilon_jK_h({{\hat B}(\hat q)}^\tau
(\mathbf{x}_i-\mathbf{x}_j)).
\end{eqnarray*}
This corresponds to the test statistics obtained from the complete
case. 

Also we can consider applying the methodology to other
testing problems such as testing for homoscedasticity, testing
for parametric quantile regression model and  testing for conditional parametric density function of  $Y$ given $\mathbf{X}$.
The relevant research is ongoing. 

In summary, the methodology is an general method that can be readily applied to many testing problems.

\section*{Appendix. Proof of the theorems}

\subsection{Conditions}\label{cond} The following conditions are assumed for
the theorems in Section 3.

\begin{itemize}

    \item [1)] $\sup E(X^2_{l}|B^\tau \mathbf{X} ) < \infty, l=1,\cdots,p$; $E(\eta^2|B^\tau \mathbf{X})<\infty$, $\sup G^2(B^\tau \mathbf{X})<\infty$ and there exists an integrable function $L(x)$ such that $|m_i(\mathbf{X},\beta,\theta)|\leq L(x)$ for all $(\beta,\theta)$ and $1\leq i\leq d+p$. $g(\mathbf{X}^\tau\beta, \theta)$ is a Borel measurable function on $R^p$ for each $\beta,\theta$ and a twice continuously differentiable real function on a compact subset of $R^p$ and $R^d$, $\Lambda$ and $\Theta$ for each $x\in R^p;$

    \item [2)]
     $nh^2\rightarrow \infty$ under the null (\ref{null}) and local alternative hypothesis (\ref{alter}); $nh^{q}\rightarrow \infty$ under global alternative hypothesis (\ref{gen-alter1}).

    \item [3)] The density $f(B^\tau \mathbf{X})$ of $B^\tau \mathbf{X}$
     on support $\mathcal{C}$ exists and
     has 2 bounded derivatives and satisfies
\begin{eqnarray*}
0< \inf_{B^\tau \mathbf{X}\in \mathcal{C}} f(B^\tau \mathbf{X}) \leq
\sup_{B^\tau \mathbf{X} \in \mathcal{C}} f(B^\tau
\mathbf{X})<\infty;
\end{eqnarray*}

    \item [4)] $K(\cdot)$ is a spherically symmetric density function with a bounded derivative and support, and
        all the moments of $K(\cdot)$ exist and $\int UU^\top K(U)dU = I$.

\item [5)] Let $\gamma=(\beta,\theta)$ and $\tilde{\gamma}_0$, the value of $\gamma$ that minimizes $\tilde{S}_{0n}(\gamma)=E[(E(Y|X)-g(\mathbf{X}^\tau\beta, \theta))^2]$, is an interior point  and is the unique minimizer of the function $\tilde{S}_{0n}$. $\Sigma_x=E(m(\mathbf{X},\beta,\theta) m(\mathbf{X},\beta,\theta)^\tau)$ is positive definite.
\end{itemize}

\begin{remark}  Condition 1) is necessary for the root-$n$ consistency of the least squares estimates $\hat\beta$ and $\hat\theta$. Condition 2)
is needed for the asymptotic normality of our statistic. In
Condition 2), $nh^2\rightarrow \infty$ is an usual assumption in
nonparametric estimation.
Conditions  3) and 4) are  commonly used in nonparametric estimation. Condition 5) is necessary for the asymptotic normality of relevant estimators.
\end{remark}

\subsection{Lemmas}\label{lem}

\begin{lemma}\label{lemma1}
Let $\mathcal{M}(t)$ be a $p\times p$ positive semi-definite matrix such that span$\{\mathcal{M}(t)\}=\mathcal{S}_{Z(t)|\mathbf{X}}.$ We can have span$\{E\{\mathcal{M}(T)\}\}$=span\{$E\{\mathcal{M}(T)\rho(T)\}\}$, here $\rho(\cdot)>0$ is some weight function.
\end{lemma}

\noindent \emph{Proof of Lemma 1}. Let $\mathbf{v}\bot $ span\{$E\{\mathcal{M}(T)\}\}$, we can have
$0=E\{\mathbf{v}^\tau\mathcal{M}(T)\mathbf{v}\}$. Due to the fact that $\mathcal{M}(t)$ is semi-definite matrix for any $t$, we can obtain that $\mathcal{M}(t)\rho(t)\mathbf{v}=0$. In other words, $\mathbf{v}\bot $ span$\{{\mathcal{M}(t)\rho(t)}\}$ for any $t$. Further, we can get $E\{\mathcal{M}(T)\rho(T)\mathbf{v}\}=0.$ Thus $\mathbf{v}\bot $ span\{$E\{\mathcal{M}(T)\rho(T)\}\}$. This follows that span\{$E\{\mathcal{M}(T)\rho(T)\}\}\subseteq$ span\{$E\{\mathcal{M}(T)\}\}.$ Another direction can be similarly shown. We conclude that span\{$E\{\mathcal{M}(T)\}\}$=span $\{E\{\mathcal{M}(T)\rho(T)\}\}$.

%

 \begin{lemma}\label{lemm2}
 Under the null hypothesis and conditions 1)-4), we  have
\begin{eqnarray*}\label{lemma2}
W_n=\frac{1}{n(n-1)}\sum_{i=1}^n\sum_{j\neq i}^n K_h({{\hat B}(\hat q)}^\tau
(\mathbf{x}_i-\mathbf{x}_j))\epsilon_iM(x_{jl})=O_p(1/\sqrt{n}),l=1,\cdots,p.
\end{eqnarray*}
where $M(\cdot)$ is continuously differentiable and
$E(M^2(X_{l})|B^\tau \mathbf{X})\leq b(B^\tau \mathbf{X})$ for
$X_{l}\in R$ and $E[b(B^\tau \mathbf{X})]<\infty$.
 \end{lemma}

\noindent \emph{Proof of Lemma \ref{lemm2}}. Under null hypothesis, $B=c\beta$ and $\hat B(\hat q)$ is consistent to $B$. For notational
convenience, denote $B_{ij}=B^\tau (\mathbf{x}_i-\mathbf{x}_j)$
and ${{\hat B}(\hat q)}_{ij}={{\hat B}(\hat q)}^\tau (\mathbf{x}_i-\mathbf{x}_j)$. Further, define
$\tilde{W}_n$ as
\begin{eqnarray*}
\tilde{W}_n&=&\frac{1}{n(n-1)}\sum_{i=1}^n\sum_{j\neq i}^n
\frac{1}{h}K(\frac{{{\hat B}(\hat q)}_{ij}}{h})
\epsilon_iM(x_{jl})=\frac{h^{\hat q}}{h}W_n.
\end{eqnarray*}
Since $\hat q\rightarrow 1$ in probability,  we
only need to show $\tilde{W}_n=O_p(1/\sqrt{n})$. Note that
\begin{eqnarray*}
\tilde{W}_n&=&\frac{1}{n(n-1)}\sum_{i=1}^n\sum_{j\neq i}^n
\frac{1}{h}K(\frac{B_{ij}}{h}) \epsilon_iM(x_{jl})\\
&&+\frac{1}{n(n-1)}\sum_{i=1}^n\sum_{j\neq i}^n
\frac{1}{h}\left(K(\frac{{{\hat B}(\hat q)}_{ij}}{h})-
K(\frac{B_{ij}}{h})\right)\epsilon_iM(x_{jl})\\
&=&W_{n1}+W_{n2}.
\end{eqnarray*}
Let $\mathbf{t}_i=(y_i,\mathbf{x}^\tau_i)^\tau$, then $W_{n1}$ can be written in a U-statistic
with the kernel as
\begin{eqnarray*}
H_n(\mathbf{t}_i,\mathbf{t}_j)=\frac{1}{2h}K(\frac{B_{ij}}{h})[\epsilon_iM(x_{jl})+\epsilon_jM(x_{il})].
\end{eqnarray*}
To apply the theory for non-degenerate U-statistic (Serfling 1980), we need to show
$E[H^2_n(\mathbf{t}_i,\mathbf{t}_j)]=o(n)$. Let $\mathbf{Z}=B^\tau
\mathbf{X}$. It can be verified  that
\begin{eqnarray*}
&&E[H^2_n(\mathbf{t}_i,\mathbf{t}_j)]\\
&\leq&2E\left[\frac{1}{2h}K(\frac{B_{ij}}{h})
\epsilon_iM(x_{jl})\right]^2+2
E\left[\frac{1}{2h}K(\frac{B_{ij}}{h})
\epsilon_jM(x_{il})\right]^2\\
&=&\int \frac{1}{h^{2}}\sigma^2({z}_i)E(M^2(x_{jl})|{z}_j)K^2(\frac{{z}_i-z_j}{h})f({z}_i)f({z}_j)
d{z}_id{z}_j\\
&\leq &\int \frac{1}{h}\sigma^2({z}_i)b({z}_i-hu)K^2(u)f({z}_i)f({z}_i-hu)d{z}_idu\\
&=&\int \frac{1}{h}\sigma^2({z})b({z})f^2({z})d{z}\cdot \int K^2(u)du+o(1/h)\\
&=&O(1/h)=o(n).
\end{eqnarray*}
Since $E(\epsilon|\mathbf{X})=0$, it can be
derived that $E(H_n(\mathbf{t}_i,\mathbf{t}_j))=0$.
 Now, consider the conditional expectation of $H_n(\mathbf{t}_i,\mathbf{t}_j)$. Also, it is easy to compute that
\begin{eqnarray*}
r_n(\mathbf{t}_i)&=&E(H_n(\mathbf{t}_i,\mathbf{t}_j)|\mathbf{t}_i)=\frac{\epsilon_i}{2h} E\left(K\big(\frac{B^\tau (\mathbf{x}_i-\mathbf{X})}{h}\big)M(X_{l})\right)\\
&=&\frac{\epsilon_i}{2h}
E\left(K\big(\frac{{z}_i-{Z}}{h}\big)E(M(X_{l})|{Z})\right)=
\frac{\epsilon_i}{2}\int E(M(X_{l})|{z}_i+hu)f({z}_i+hu)K(u)du\\
&=&\frac{\epsilon_i f({z}_i)
E(M(X_{l})|{z}_i)}{2}+l_n(\mathbf{t}_i).
\end{eqnarray*}
Denote $\hat {W}_n$ as the ``projection" of the statistic $W_{n1}$ as: \begin{eqnarray*} \sqrt{n} \hat
W_n&=&\frac{2}{\sqrt{n}}\sum_{i=1}^n
r_n(\mathbf{t}_i)=\frac{1}{\sqrt{n}}\sum_{i=1}^n \epsilon_i f({z}_i)
E(M(X_{l})|{z}_i)+\frac{2}{\sqrt{n}}\sum_{i=1}^n l_n(\mathbf{t}_i)\\
&=&O_p(1).
 \end{eqnarray*}
 The last equation follows from the fact that $E(l^2_n(\mathbf{t}_i))=O(h^2)\rightarrow 0$ due to the Lipschitz condition for the function $E(M({X}_{l})|\cdot)f(\cdot)$.
As a result, we have $W_{n1}=O_p(\hat W_n)=O_p(1/\sqrt{n})$. Denote
$$W^*_{n2}=\frac{1}{n(n-1)}\sum_{i=1}^n\sum_{j\neq i}^n
\frac{1}{h}K'(\frac{B_{ij}}{h})(\mathbf{x}_i-\mathbf{x}_j)^\tau
\epsilon_iM(X_{jl})\times\frac{{{\hat B}(\hat q)}-B}{h}.$$ Then for the term $W_{n2}$,
we  have
\begin{eqnarray*}
W_{n2}&=&W^*_{n2}+o_p(W^*_{n2}).
\end{eqnarray*}
Since $K(\cdot)$ is  spherically symmetric, similar to $W_{n1}$, the following term $$\frac{1}{n(n-1)}\sum_{i=1}^n\sum_{j\neq i}^n \frac{1}{h}K'(\frac{B_{ij}}{h})(\mathbf{x}_i-\mathbf{x}_j)^\tau
\epsilon_iM(X_{jl})$$ can be rewritten as a U-statistic. Then we can similarly show that this term is also of order $O_p(1/\sqrt{n})$.
As  $||{{\hat B}(\hat q)}-B||_2=O_p(1/\sqrt{n})$, and under the
condition $1/nh^2\rightarrow 0$, we can
obtain that $W_{n2}=o_p(1/\sqrt{n})$. Thus we can conclude that
$W_n=O_p(1/\sqrt{n}).$ The proof is completed.  \hfill $\fbox{}$

\bigskip
Before we establish the asymptotic theory of our statistic under the null and local alternatives,
we develop the following lemma about the asymptotic property of $\hat\beta$ and $\hat\theta$. This is necessary because it is defined under the null hypothesis.

\begin{lemma}\label{lemm3}
 Under the local alternative and conditions 1), 5), we  have
\begin{eqnarray*}
\sqrt{n}(\hat\gamma-\gamma)&=&\Sigma^{-1}_x\frac{1}{\sqrt{n}}\sum_{i=1}^n m(\mathbf{x}_i,\beta,\theta)\eta_i +\Sigma^{-1}_xC_n\sqrt{n}E(m(\mathbf{X},\beta,\theta)G(B^\tau \mathbf{X}))\\
&&+(\tilde{\Sigma}^{-1}-\Sigma^{-1}_x)C_n\sqrt{n}E(m(\mathbf{X},\beta,\theta)G(B^\tau \mathbf{X}))+ o_p(1).
\end{eqnarray*}
Here $\tilde{\Sigma}=n^{-1}\sum_{i=1}^n
m(\mathbf{x}_i,\beta,\theta)m(\mathbf{x}_i,\beta,\theta)^\tau$ and $\gamma=(\beta,\theta)^\tau$.
 \end{lemma}

\noindent \emph{Proof of Lemma~\ref{lemm3}.}  Under the regularity conditions designed in Jennrich (1969), similar
to the derivation of Theorems~ 6 and 7 in Jennrich (1969),
$\hat\gamma$ is a strongly consistent estimate of $\gamma$. If we let
$E_n=\tilde{\Sigma}^{-1}n^{-1}\sum_{i=1}^n m(\mathbf{x}_i,\beta,\theta)(y_i-g(\beta^\tau\mathbf{x}_i,\theta))$, we can
further have:
\begin{eqnarray}\label{beta}
\hat\gamma-\gamma&=&\tilde{\Sigma}^{-1}\frac{1}{n}\sum_{i=1}^n m(\mathbf{x}_i,\beta,\theta)(y_i-g(\beta^\tau\mathbf{x}_i,\theta))+o_p(E_n) \nonumber\\
&=&(\tilde{\Sigma}^{-1}-\Sigma^{-1}_x)\frac{1}{n}\sum_{i=1}^n
m(\mathbf{x}_i,\beta,\theta)\eta_i \nonumber\\
&&+(\tilde{\Sigma}^{-1}-\Sigma^{-1}_x)\frac{1}{n}\sum_{i=1}^n
m(\mathbf{x}_i,\beta,\theta)C_nG(B^\tau \mathbf{x}_i) \nonumber\\
&&+\Sigma^{-1}_x\frac{1}{n}\sum_{i=1}^n m(\mathbf{x}_i,\beta,\theta)\eta_i \nonumber\\
&&+\Sigma^{-1}_x\frac{1}{n}\sum_{i=1}^n m(\mathbf{x}_i,\beta,\theta)C_nG(B^\tau \mathbf{x}_i)+o_p(E_n) \nonumber\\
&=:&\sum_{i=1}^4 I_{ni}+o_p(E_n).
\end{eqnarray}

Due to the consistency of $\tilde{\Sigma}$ for $\Sigma_x$, we can easily conclude that
\begin{eqnarray*}
\sqrt{n}(\hat\gamma-\gamma)&=&\Sigma^{-1}_x\frac{1}{\sqrt{n}}\sum_{i=1}^n m(\mathbf{x}_i,\beta,\theta)\eta_i +\Sigma^{-1}_xC_n\sqrt{n}E(m(\mathbf{X},\beta,\theta)G(B^\tau \mathbf{X}))\\
&&+(\tilde{\Sigma}^{-1}-\Sigma^{-1}_x)C_n\sqrt{n}E(m(\mathbf{X},\beta,\theta)G(B^\tau \mathbf{X}))+ o_p(1).
\end{eqnarray*}

\vspace{3mm}

\subsection{Proofs of the theorems}

\noindent \emph{Proof of Theorem 1.} First, $V_n$ can be  decomposed
as, noting the symmetry of $K_h(\cdot)$,
\begin{eqnarray}\label{Tnd}
V_n&=&\frac{1}{n(n-1)}\sum_{i=1}^n\sum_{j\neq i}^n K_h({{\hat B}(\hat q)}_{ij})\epsilon_i\epsilon_j\nonumber\\
&&-\frac{2}{n(n-1)}\sum_{i=1}^n\sum_{j\neq i}^n K_h({{\hat B}(\hat q)}_{ij})\epsilon_i m(\mathbf{x}_j,\beta,\theta)^\tau(\hat\gamma-\gamma)\nonumber \\
&&+ (\hat\gamma-\gamma)^{\tau}\frac{1}{n(n-1)}\sum_{i=1}^n\sum_{j\neq i}^n K_h({{\hat B}(\hat q)}_{ij})m(\mathbf{x}_i,\beta,\theta)m(\mathbf{x}_j,\beta,\theta)^\tau(\hat\gamma-\gamma)\nonumber\\
&&+o_p(V^*_n)\nonumber \\
&=:&V_{n1}-V_{n2}+V_{n3}+o_p(V^*_n),
\end{eqnarray}
where $V^*_n$ denotes the term $V_{n1}-V_{n2}+V_{n3}.$

Consider the term $V_{n2}$. Under the conditions designed for
Theorem~\ref{the1}, and from Lemmas~\ref{lemm2} and \ref{lemm3}, we can get that
$V_{n2}=O_p(1/n)$. This yields that $nh^{1/2}V_{n2}=o_p(1).$

Now we deal with the term $V_{n3}$. Rewrite it as
\begin{eqnarray*}
V_{n3}&=&(\hat\gamma-\gamma)^\tau\cdot\frac{1}{n(n-1)}\sum_{i=1}^n\sum_{j\neq i}^n K_h({{\hat B}(\hat q)}_{ij})m(\mathbf{x}_i,\beta,\theta)m(\mathbf{x}_j,\beta,\theta)^\tau\cdot(\hat\gamma-\gamma).
\end{eqnarray*}
Under null hypothesis, $B=c\beta$ and $\hat B(\hat q)$ is consistent to $B$. A similar argument for proving Lemma \ref{lemm2} can be used to
derive that
\begin{eqnarray*}
&&\frac{1}{n(n-1)}\sum_{i=1}^n\sum_{j\neq i}^n K_h({{\hat B}(\hat q)}_{ij})m(\mathbf{x}_i,\beta,\theta)m(\mathbf{x}_j,\beta,\theta)^\tau\\
&=&E(m(\mathbf{X},\beta,\theta)m(\mathbf{X},\beta,\theta)^\tau f(\beta^\tau
\mathbf{X}))+o_p(1).
\end{eqnarray*}
By the rate of $\hat \gamma -\gamma$, $V_{n3}=O_p(1/n)$.
Consequently,  $nh^{1/2}V_{n3}=o_p(1).$

Finally, deal with the term $V_{n1}$. Note that  we have the
following decomposition:
 \begin{eqnarray*}
 V_{n1}&=&\frac{1}{n(n-1)}\sum_{i=1}^n\sum_{j\neq i}^n K_h(B_{ij})\epsilon_i\epsilon_j\\
 &&+\frac{1}{n(n-1)}\sum_{i=1}^n\sum_{j\neq i}^n (K_h({{\hat B}(\hat q)}_{ij})-K_h(B_{ij}))\epsilon_i\epsilon_j\\
 &=:&V_{n1,1}+V_{n1,2}.
 \end{eqnarray*}
 For the term $V_{n1,1}$, since in this paper, we always assume that the dimension of $B^\tau \mathbf{X}$ is fixed, it is an U-statistic. Note that under the null hypothesis, $q=1$ and $\hat q\rightarrow 1$. It is easy to derive the asymptotic normality: $nh^{1/2}V_{n1,1}\rightarrow N(0,{Var})$. Here
 \begin{eqnarray*}
 {Var}=2\int K^2(u)du\cdot \int (\sigma^2(\mathbf{z}))^2f^2(\mathbf{z})d\mathbf{z}
 \end{eqnarray*}
with  $\mathbf{Z}=B^\tau
\mathbf{X},\sigma^2(\mathbf{z})=E(\epsilon^2|\mathbf{Z}=\mathbf{z})$. See a similar argument as that for Lemma 3.3 a) in Zheng (1996). We then omit the details.

 Denote $$V^*_{n1,2}=\frac{h}{h^{\hat q}}\cdot\frac{1}{n(n-1)}\sum_{i=1}^n\sum_{j\neq i}^n
\frac{1}{h}K'(\frac{B_{ij}}{h})(\mathbf{x}_i-\mathbf{x}_j)^\tau\epsilon_i\epsilon_j\cdot \frac{{{\hat B}(\hat q)}-B}{h}.$$ An application of Taylor
expansion yields
 $$V_{n1,2}=V^*_{n1,2}+o_p(V^*_{n1,2}).$$
 Since $K(\cdot)$ is spherical symmetric, the term $$\frac{1}{n(n-1)}\sum_{i=1}^n\sum_{j\neq i}^n
\frac{1}{h}K'(\frac{B_{ij}}{h})(\mathbf{x}_i-\mathbf{x}_j)^\tau\epsilon_i\epsilon_j$$ can be considered as an U-statistic. Further note that
\begin{eqnarray*}
&&E\left(K'(\frac{B_{ij}}{h})(\mathbf{x}_i-\mathbf{x}_j)^\tau\epsilon_i\epsilon_j|
\mathbf{x}_i,y_i\right)\\
&=&E\left(E\{K'(\frac{B_{ij}}{h})(\mathbf{x}_i-\mathbf{x}_j)^\tau\epsilon_i\epsilon_j|
\mathbf{x}_i,y_i,\mathbf{x}_j\}|\mathbf{x}_i,y_i\right)\\
&=&E\left(K'(\frac{B_{ij}}{h})(\mathbf{x}_i-\mathbf{x}_j)^\tau \epsilon_iE\{\epsilon_j|\mathbf{x}_j\}|\mathbf{x}_i,y_i\right)=0.
\end{eqnarray*}
Thus the above U-statistic is degenerate.
 Using a similar argument for the term $V_{n1,1}$ above,
together with  $||{{\hat B}(\hat q)}-B||_2=O_p(1/\sqrt{n})$ and
$1/nh^2\rightarrow 0$, it results in that
$nh^{1/2}V^*_{n1,2}=o_p(1)$. Thus we can have
$nh^{1/2}V_{n1}\rightarrow N(0,{Var})$.

 Combining the above results for the terms $V_{ni},i=1,2,3$, we conclude that
 \begin{eqnarray*}nh^{1/2}V_{n}\Rightarrow N(0,{Var}).\end{eqnarray*}

Since ${Var}$ is actually unknown, an estimate is defined as
$$\widehat{Var}=\frac{2}{n(n-1)}\sum_{i=1}^n\sum_{j\neq i}^n
\frac{1}{h^{\hat q}}K^2(\frac{{{\hat B}(\hat q)}^\tau
(\mathbf{x}_i-\mathbf{x}_j)}{h})\hat\epsilon^2_i\hat \epsilon^2_j.$$

As the proof is rather straightforward, we then only give a very
brief description. Since $\hat\beta$ is consistent under the null
hypothesis, some elementary computations lead to an asymptotic
presentation:
$$\widehat{Var}=\frac{2}{n(n-1)}\sum_{i=1}^n\sum_{j\neq i}^n
\frac{1}{h^{\hat q}}K^2(\frac{{{\hat B}(\hat q)}^\tau
(\mathbf{x}_i-\mathbf{x}_j)}{h})\epsilon^2_i\epsilon^2_j+o_p(1).$$
Using a similar argument as that for Lemma~\ref{lemm2}, we get
$$\widehat{Var}=\frac{2}{n(n-1)}\sum_{i=1}^n\sum_{j\neq i}^n
\frac{1}{h^{\hat q}}K^2(\frac{B^\tau
(\mathbf{x}_i-\mathbf{x}_j)}{h})\epsilon^2_i\epsilon^2_j+o_p(1).$$
The consistency will be derived by using U-statistic theory. The
proof is finished. \hfill $\fbox{}$

\bigskip

\vspace{2mm}

\noindent \emph{Proof of Theorem \ref{theo2}}.
We first investigate the $\hat q$ determined by the BIC criterion for DEE. We also adopt the same conditions as those in Theorem 4 of Zhu et al. (2010a).

 From the argument used for proving Theorem~3.2 of Li et al. (2008), we can know that to obtain that $\mathcal{M}_{n,n}-\mathcal{M}=O_p(C_n)$, we only need to show that $\mathcal{M}_{n}(t)-\mathcal{M}(t)=O_p(C_n)$ uniformly. Now we investigate the term $\mathcal{M}_n(t)$ for every $t$.
Here $\mathcal{M}(t)=\Sigma^{-1}Var(E(\mathbf{X}|Z(t)))=\Sigma^{-1}(\mu_1-\mu_0)(\mu_1-\mu_0)^\tau p(1-p)$ here $\Sigma$ is the covariance matrix of $\mathbf{X}$, $\mu_j=E(\mathbf{X}|Z(t)=j)$ with $j=0$ and 1 and $p=E(I(Y\leq t))$.
Further note that
\begin{eqnarray*}
\mu_1-\mu_0&=&\frac{E(\mathbf{X}I(Y\leq t))}{p}-\frac{E(\mathbf{X}I(Y>t))}{1-p}\\
&=&\frac{E(\mathbf{X}I(Y\leq t))-E(\mathbf{X})E(I(Y\leq t))}{p(1-p)}.
\end{eqnarray*}
Thus from  Lemma~\ref{lemma1},  $\mathcal{M}(t)$ can also be taken to be $$\mathcal{M}(t)=\Sigma^{-1}\Big[E\{(\mathbf{X}-E(\mathbf{X}))I(Y\leq t)\}\Big]\Big[E\{(\mathbf{X}-E(\mathbf{X}))I(Y\leq t)\}\Big]^\tau.$$
For ease of illustration, we denote $m(t)=E\{(\mathbf{X}-E(\mathbf{X}))I(Y\leq t)\}$, then $\mathcal{M}(t)=\Sigma^{-1}m(t)m(t)^\tau=\Sigma^{-1}L(t)$.  $m(t)$ is estimated by
$$m_n(t)=:n^{-1}\sum_{i=1}^n(\mathbf{x}_i-\overline{\mathbf{x}})I(y_i\leq t).$$
Thus, $\mathcal{M}_n(t)$ can be taken to be $$\mathcal{M}_n(t)=\hat\Sigma^{-1}m_n(t)m^\tau_n(t)=\hat\Sigma^{-1}L_n(t).$$
Denote the response under the null and local alternative as $Y$ and $Y_n$ respectively.
Note that
\begin{eqnarray*}
&&\frac{1}{n}\sum_{i=1}^n\mathbf{x}_iI(y_{in}\leq t)-E(\mathbf{X}I(Y\leq t))\\
&=&\frac{1}{n}\sum_{i=1}^n[\mathbf{x}_iI(y_{in}\leq t)-E(\mathbf{X}I(Y_n\leq t))]+E(\mathbf{X}I(Y_n\leq t))-E(\mathbf{X}I(Y\leq t)).
\end{eqnarray*}
By the Lindeberg-Levy central limit theorem, the first term has the rate of  order $O_p(n^{-1/2})$.
Now we consider the second term. Note that
\begin{eqnarray*}
E(\mathbf{X}I(Y_n\leq t))-E(\mathbf{X}I(Y\leq t))&=&E\Big(\mathbf{X}[P(Y_n\leq t|\mathbf{X})-P(Y\leq t|\mathbf{X})]\Big).
\end{eqnarray*}
Recall that $Y_n=Y+ C_n G(B^\tau \mathbf{X})$ and denote the conditional density and  distribution function of $Y$ given $\mathbf{X}$ as $f_{Y|\mathbf{X}}(\cdot)$ and $F_{Y|\mathbf{X}}(\cdot)$ respectively. We can then have
\begin{eqnarray*}
P(Y_n\leq t|\mathbf{X})-P(Y\leq t|\mathbf{X})&=&F_{Y|\mathbf{X}}(t-C_nG(B^\tau \mathbf{X}))-F_{Y|\mathbf{X}}(t)\\
&=&-C_nG(B^\tau \mathbf{X})f_{Y|\mathbf{X}}(t)+O_p(C_n).
\end{eqnarray*}
From this, we can conclude that $n^{-1}\sum_{i=1}^n\mathbf{x}_iI(y_{in}\leq t)-E(\mathbf{X}I(Y\leq t))=O_p(C_n)$. Similarly,  $m_n(t)-m(t)=O_p(C_n), L_n(t)-L(t)=O_p(C_n)$ and $\mathcal{M}_n(t)-\mathcal{M}(t)=O_p(C_n)$. Finally, similar to the argument used for proving Theorem 3.2 of Li et al. (2008),  $\mathcal{M}_{n,n}-\mathcal{M}=O_p(C_n)$.
Now we turn to prove the consistency of BIC criterion-based estimate when DEE is used. By the definition in Section~2, for $l>1$, we  have

\begin{eqnarray*}
G(1)-G(l)=D_n\frac{l(l+1)-2}{p}-\frac{n\sum_{i=2}^l\log{(\hat\lambda_i+1)}-\hat\lambda_i}
{2\sum_{i=1}^p\log{(\hat\lambda_i+1)}-\hat\lambda_i}.
\end{eqnarray*}
Invoking $\mathcal{M}_{n,n}-\mathcal{M}=O_p(C_n)$, $\hat\lambda_i-\lambda_i=O_p(C_n)$. Note that $\log{(\hat\lambda_i+1)}-\hat\lambda_i=-\hat\lambda^2_i+o_p(\hat\lambda^2_i)$
and $\lambda_i=0$ for any $i>1$. We can obtain that $\sum_{i=2}^l\log{(\hat\lambda_i+1)}-\hat\lambda_i=O_p(C^2_n)$ and $\sum_{i=1}^p\log{(\hat\lambda_i+1)}-\hat\lambda_i\rightarrow b$ in probability for some $b<0$.
Taking $D_n=n^{1/2}$ and $C_n=n^{-1/2}h^{-1/4}$, it is easy to see that
$$\frac{nC^2_n}{D_n}=(nh)^{-1/2}\rightarrow 0.$$
Since $l(l+1)>2$ for any $l>1$, $P(G(1)>G(l))\rightarrow 1.$ In other words,$P(\hat q=1)\rightarrow 1$.

\vspace{4mm} From Zhu et al. (2010a), the matrix $\mathcal{M}$ that is based on either SIR or SAVE can satisfy the consistency $\mathcal{M}_{n,n}$ requires. Thus, the estimate $\hat q$ that is based on $DEE_{SIR}$ or $DEE_{SAVE}$ can have the above consistency. We then omit the detail here.

\vspace{4mm}

We now turn to consider the $\hat q$ being selected by the BIC criterion with MAVE. Assume the  same conditions in Wang and Yin (2008), and for simplicity assume that $\mathbf{X}$
has a compact support over which its density is positive. Recall the
definition of $B(k)$ for any $k$, we have that
$Y=E(Y|B^\tau(k)\mathbf{X})+\epsilon$, where
$E(\epsilon|B^\tau(k)\mathbf{X})=0$. Suppose that the orthogonal
$p\times k$ matrix ${{\hat B}(k)}$ is the sample estimate of $B(k)$.

Note that under the local alternative hypothesis (\ref{alter}), we can have
\begin{eqnarray*}
Y&=&g(\beta^\tau \mathbf{X},\theta)+C_n G(B^\tau \mathbf{X})+\eta;\\
E(Y|B^\tau(k)\mathbf{X})&=&g(\beta^\tau \mathbf{X},\theta)+ C_n E(G(B^\tau \mathbf{X})|B^\tau(k)\mathbf{X}).
\end{eqnarray*}
Similar to the argument used in Wang and Yin (2008), under the local alternatives, we have
\begin{eqnarray*}
\frac{1}{n}RSS_k&=&E\{Y-E(Y|B^\tau(k)\mathbf{X})\}^2+O_p(\frac{1}{\sqrt{n}}+\frac{1}{nh^k})\\
&=&E\{\eta^2\}+O_p(C^2_n)+O_p(\frac{1}{\sqrt{n}}+\frac{1}{nh^k}).
\end{eqnarray*}
The second equation is based on the fact that
$E\{\eta[G(B^\tau \mathbf{X})-E(G(B^\tau \mathbf{X})|B^\tau(k)\mathbf{X})]\}=0$ by using
$E(\eta|\mathbf{X})=0$.
Consequently, the following results hold:
\begin{eqnarray*}
\frac{RSS_k-RSS_1}{n}&=&O_p(C^2_n)+O_p(\frac{1}{\sqrt{n}}+\frac{1}{nh^k}).
\end{eqnarray*}
Recall that the BIC criterion is
$BIC_k=\log(RSS_k/n)+\log(n)k/\min(nh^k,\sqrt{n})$.
Thus we can obtain that
$$BIC_k-BIC_1=\log(RSS_k/RSS_1)+\frac{\log(n)k}{\min(nh^k_{k,opt},\sqrt{n})}
-\frac{\log(n)}{\min(nh_{1,opt},\sqrt{n})}.$$
Consider the term $\log(RSS_k/RSS_1)$ first. We can easily get
\begin{eqnarray*}
\log(RSS_k/RSS_1)&=&\log(1+\frac{RSS_k-RSS_1}{RSS_1})\\
&=&\frac{RSS_k-RSS_1}{RSS_1}+o_p(\frac{RSS_k-RSS_1}{RSS_1})\\
&=&O_p(C^2_n)+O_p(\frac{1}{\sqrt{n}}+\frac{1}{nh^k}).
\end{eqnarray*}
Take $h_{k,opt}=O(n^{-1/(4+k)})$ for each $k\geq 1$.
Note that for $k<4$, $\sqrt{n}=o(nh^k_{k,opt})$.
Then we can get, for large $n$,
$$
\sqrt{n}\left[\frac{\log(n)k}{\min(nh^k_{k,opt},\sqrt{n})}
-\frac{\log(n)}{\min(nh_{1,opt},\sqrt{n})}\right]=\log(n)(k-1).$$
Also note that $\sqrt{n}C^2_n=o(1)$ where $C_n=n^{-1/2}h^{-1/4}$. Thus for $k=2$ and 3, in probability,
$$\sqrt{n}(BIC_k-BIC_1)\to \infty.$$
For $k>4$, $nh^k_{k,opt}=o(\sqrt{n})$.
Then
$$
nh^k_{k,opt}\left[\frac{\log(n)k}{\min(nh^k_{k,opt},\sqrt{n})}
-\frac{\log(n)}{\min(nh_{1,opt},\sqrt{n})}\right]
=nh^k_{k,opt}\left[\frac{\log(n)k}{nh^k_{k,opt}}
-\frac{\log(n)}{\sqrt{n}}\right]\Rightarrow\infty.$$
Note that $nh^k_{k,opt}C^2_n=(nh^{2k}_{k,opt})^{1/2}\cdot \sqrt{n}C^2_n=o(1)$ with $C_n=n^{-1/2}h^{-1/4}$. Then in probability
we have $$nh^k_{k,opt}(BIC_k-BIC_1)\to \infty,$$ for $k>4$.
Finally, consider $k=4$. In this situation, $nh^k_{k,opt}=\sqrt{n}$ and thus
$$
\sqrt{n}\left[\frac{\log(n)k}{\min(nh^k_{k,opt},\sqrt{n})}
-\frac{\log(n)}{\min(nh_{1,opt},\sqrt{n})}\right]=3\log(n).$$
Then, it is easy to see that in probability  $$\sqrt{n}(BIC_4-BIC_1)\to \infty.$$
Hence, for large $n$, $\hat q\rightarrow 1$. We finally conclude that
$P(BIC_k>BIC_1)\to 1$ for any $k>1$.
In other words, $P(\hat q=1)\rightarrow 1$. \hfill $\fbox{}$

\

\

\noindent \emph{Proof of Theorem 3.}
We first consider the global alternative (\ref{gen-alter1}).  Under this  alternative, Proposition~\ref{prop1} shows that $\hat q\rightarrow q\geq 1$.
According to White (1981),  $\hat\gamma$ is  a root-$n$ consistent estimator of $\tilde{\gamma}_0$ which is different from the true value $\gamma$ under the null hypothesis. Let $\Delta(\mathbf{x}_i)=
G(B^\tau \mathbf{x}_i)-g(\tilde{\beta}^\tau_0 \mathbf{x}_i,\tilde\theta_0)$. Then
$\hat\epsilon_i=\eta_i+\Delta(\mathbf{x}_i)-(g(\hat\beta^\tau \mathbf{x}_i,\hat\theta)-g(\tilde{\beta}^\tau_0 \mathbf{x}_i,\tilde\theta_0)$.
It is then easy to see that $V_n\Rightarrow E(\Delta^2(\mathbf{X})f(B^\tau \mathbf{X}))$ by using the U-statistics theory. Similarly, we can also have that in probability $\widehat{Var}$ converges to a positive value which is different from  ${Var}$. Then obviously, we can obtain that $T_n/(nh^{1/2})\to Constant>0$ in probability.

Under the local alternatives (\ref{alter}),
$V_n$ has the following decomposition by Taylor expansion:
\begin{eqnarray}\label{Tnd}
V_n&=&\frac{1}{n(n-1)}\sum_{i=1}^n\sum_{j\neq i}^n K_h({{\hat B}(\hat q)}^\tau (\mathbf{x}_i-\mathbf{x}_j))(\eta_i+C_nG(B^\tau \mathbf{x}_i))(\eta_j+C_n G(B^\tau\mathbf{x}_j))\nonumber\\
&&-\frac{2}{n(n-1)}\sum_{i=1}^n\sum_{j\neq i}^n K_h({{\hat B}(\hat q)}^\tau (\mathbf{x}_i-\mathbf{x}_j))(\eta_i+C_nG(B^\tau \mathbf{x}_i)) m(\mathbf{x}_j,\beta,\theta)^\tau(\hat\gamma-\gamma)\nonumber \\
&&+ \frac{1}{n(n-1)}\sum_{i=1}^n\sum_{j\neq i}^n K_h({{\hat B}(\hat q)}^\tau (\mathbf{x}_i-\mathbf{x}_j))m(\mathbf{x}_i,\beta,\theta)^\tau(\hat\gamma-\gamma)
m(\mathbf{x}_j,\beta,\theta)^\tau(\hat\gamma-\gamma)\nonumber \\
&&+o_p(\overline{V}^*_n)\nonumber \\
&=:&\overline{V}_{n1}-\overline{V}_{n2}+\overline{V}_{n3}+o_p(\overline{V}^*_n)
\end{eqnarray}
where  $\overline{V}^*_n=
\overline{V}_{n1}-\overline{V}_{n2}+\overline{V}_{n3}.$

For the term $\overline{V}_{n2}$ in (\ref{Tnd}), it follows that
\begin{eqnarray*}
\overline{V}_{n2}&=&\frac{2}{n(n-1)}\sum_{i=1}^n\sum_{j\neq i}^n K_h({{\hat B}(\hat q)}^\tau (\mathbf{x}_i-\mathbf{x}_j))\eta_im(\mathbf{x}_j,\beta,\theta)^\tau(\hat\gamma-\gamma)\\
&&+C_n\frac{2}{n(n-1)}\sum_{i=1}^n\sum_{j\neq i}^n K_h({{\hat B}(\hat q)}^\tau (\mathbf{x}_i-\mathbf{x}_j))G(B^\tau \mathbf{x}_i)m(\mathbf{x}_j,\beta,\theta)^\tau(\hat\gamma-\gamma)\\
&=&\overline{V}_{n2,1}(\hat\gamma-\gamma)+C_n\overline{V}^\tau_{n2,2}(\hat\gamma-\gamma).
\end{eqnarray*}
From Lemma~\ref{lemm2}, we have
$\overline{V}_{n2,1}=O_p(n^{-1/2})$. It can also be proved that
 \begin{eqnarray*}
 \overline{V}_{n2,2}&=&E(G(B^\tau \mathbf{X})m(\mathbf{X},\beta,\theta)f(\beta^\tau \mathbf{X}))+o_p(1).
\end{eqnarray*}
When  $C_n=n^{-1/2}h^{-1/4}$, Lemma~\ref{lemm3} implies that
\begin{eqnarray}\label{barVn2}
&& nh^{1/2}\overline{V}_{n2}\nonumber\\
&=&nh^{1/2}\Big[O_p(n^{-1/2})O_p(C_n) \nonumber\\
&&+
2C^2_n E^\tau(G(B^\tau \mathbf{X})m(\mathbf{X},\beta,\theta)f(\beta^\tau \mathbf{X}))\Sigma^{-1}_xE(G(B^\tau \mathbf{X})m(\mathbf{X},\beta,\theta))\Big]\nonumber\\
&=&2E^\tau(G(B^\tau \mathbf{X})m(\mathbf{X},\beta,\theta)f(\beta^\tau \mathbf{X}))\Sigma^{-1}_xE(G(B^\tau \mathbf{X})m(\mathbf{X},\beta,\theta))+o_p(1). \nonumber\\
\end{eqnarray}
Now, we turn to consider the term $\overline{V}_{n3}$. It is easy to see that
\begin{eqnarray*}
\overline{V}_{n3}&=&(\hat\gamma-\gamma)^\tau E[m(\mathbf{X},\beta,\theta)m(\mathbf{X},\beta,\theta)^\tau f(\beta^\tau \mathbf{X})](\hat\gamma-\gamma)+o_p(C^2_n)\\
&=&C^2_nE^\tau[G(B^\tau \mathbf{X})m(\mathbf{X},\beta,\theta)]\Sigma^{-1}_x E[m(\mathbf{X},\beta,\theta)m(\mathbf{X},\beta,\theta)^\tau f(\beta^\tau \mathbf{X})] \\
&& \times \Sigma^{-1}_x E[G(B^\tau \mathbf{X})m(\mathbf{X},\beta,\theta)]+o_p(C^2_n).
\end{eqnarray*}
As a result, when $C_n=n^{-1/2}h^{-1/4}$ , we  obtain
\begin{eqnarray}\label{barVn3}
nh^{1/2} \overline{V}_{n3}&=&E^\tau[G(B^\tau \mathbf{X})m(\mathbf{X},\beta,\theta)]\Sigma^{-1}_x E[m(\mathbf{X},\beta,\theta)m(\mathbf{X},\beta,\theta)^\tau f(\beta^\tau \mathbf{X})]\nonumber \\
&& \times \Sigma^{-1}_x E[G(B^\tau \mathbf{X})m(\mathbf{X},\beta,\theta)]+o_p(1).
\end{eqnarray}
Now we investigate the term $\overline{V}_{n1}$ in (\ref{Tnd}), it
can be decomposed as:
\begin{eqnarray*}
\overline{V}_{n1}&=&\frac{1}{n(n-1)}\sum_{i=1}^n\sum_{j\neq i}^n K_h({{\hat B}(\hat q)}^\tau(\mathbf{x}_i-\mathbf{x}_j))\eta_i\eta_j\\
&&+C_n\frac{2}{n(n-1)}\sum_{i=1}^n\sum_{j\neq i}^n K_h({{\hat B}(\hat q)}^\tau(\mathbf{x}_i-\mathbf{x}_j))\eta_i G(B^\tau \mathbf{x}_j)\\
&&+C^2_n\frac{1}{n(n-1)}\sum_{i=1}^n\sum_{j\neq i}^n K_h({{\hat B}(\hat q)}^\tau(\mathbf{x}_i-\mathbf{x}_j))G(B^\tau \mathbf{x}_i)G(B^\tau \mathbf{x}_j)\\
&=&\overline{V}_{n1,1}+
C_n\overline{V}_{n1,2}+C^2_n\overline{V}_{n1,3}.
\end{eqnarray*}
From the proof of Theorem~\ref{the1} and the conclusion of Lemma~\ref{lemm2}, we  have
\begin{eqnarray*}
&&\overline{V}_{n1,2}=O_p(n^{-1/2}),\\
&&\overline{V}_{n1,3}=E(G^2(B^\tau \mathbf{X})f(\beta^\tau
\mathbf{X}))+o_p(1).
\end{eqnarray*}
Note that $nh^{1/2}\overline{V}_{n1,1}\Rightarrow N(0,Var)$.
Consequently, when $C_n=n^{-1/2}h^{-1/4}$
\begin{eqnarray}\label{barVn1}
nh^{1/2}\overline{V}_{n1}\Rightarrow N(E(G^2(B^\tau
\mathbf{X})f(\beta^\tau \mathbf{X})), {Var}).
\end{eqnarray}
Combining equations (\ref{barVn2}), (\ref{barVn3}) and
(\ref{barVn1}), we can have
\begin{eqnarray*}
nh^{1/2}V_n\Rightarrow N(\mu, {Var}),
\end{eqnarray*}
where \begin{eqnarray*}
\mu&=&E[G^2(B^\tau \mathbf{X})f(\beta^\tau \mathbf{X})]-2E^\tau[H(\mathbf{X})f(\beta^\tau \mathbf{X})]\Sigma^{-1}_xE[H(\mathbf{X})]\\
&&+E^\tau[H(\mathbf{X})]\Sigma^{-1}_x
E[m(\mathbf{X},\beta,\theta)m(\mathbf{X},\beta,\theta)^\tau f(\beta^\tau
\mathbf{X})]\Sigma^{-1}_x E[H(\mathbf{X})]\\
&=&E\left[\Big(G(B^\tau \mathbf{X})-m(\mathbf{X},\beta,\theta)^\tau\Sigma^{-1}_xE[H(\mathbf{X})]\Big)^2f(\beta^\tau \mathbf{X})\right].
\end{eqnarray*}
here $H(\mathbf{X})=G(B^\tau
\mathbf{X})m(\mathbf{X},\beta,\theta)$.

Let $T_n$ be the standardized version of $V_n$ of the form:
\begin{eqnarray*}
T_n&=&\frac{h^{(1-\hat q)/2}\sum_{i=1}^n\sum_{j\neq i}^n \hat \epsilon_i\hat
\epsilon_jK(\frac{{{\hat B}(\hat q)}^\tau
(\mathbf{x}_i-\mathbf{x}_j)}{h})}{\{2\sum_{i=1}^n\sum_{j\neq i}^n
K^2(\frac{{{\hat B}(\hat q)}^\tau
(\mathbf{x}_i-\mathbf{x}_j)}{h})\hat\epsilon^2_i\hat
\epsilon^2_j\}^{1/2}}.
\end{eqnarray*}
Note that when $C_n=n^{-1/2}h^{-1/4}$, $\widehat{Var}$ is still a
consistent estimate of ${Var}$.  From Lemma~\ref{lemm3}, we can easily have
that $\hat\beta$ is also consistent under the local alternative.
Thus both $\hat\beta$ and $\widehat{Var}$ are still consistent to
$\beta$ and ${Var}$ under the local alternatives.
 Thus $T^2_n\Rightarrow \chi^2_1(\mu^2/{Var}).$

 When $C_n$ has a slower convergence rate than $n^{-1/2}h^{-1/4}$, the above arguments  can show that the test statistic goes to infinity in probability. The details are omitted. Theorem~\ref{theo3} is proved. \hfill $\fbox{}$

\clearpage

\newpage

\begin{table}
\scriptsize{ \caption{Empirical sizes and powers of $\widetilde{T}^{MAVE}_n$ and $T^{DEE}_n$ for $H_0$ vs.
$H_{11}$ and $H_{12}$, with $X\sim N(0,\Sigma_i)$, $i=1,2$ and $\epsilon\sim
N(0,1)$.}
 \label{tab1}
\begin{center}
\scriptsize
\begin{tabular}{cccccccccccccc}
\hline
 & $a$  & \multicolumn{4}{c}{$\widetilde{T}^{MAVE}_n$} & &  \multicolumn{4}{c}{$T^{DEE}_n$}\\
 &&\multicolumn{2}{c}{$n=50$} &  \multicolumn{2}{c}{$n=100$} && \multicolumn{2}{c}{$n=50$} &\multicolumn{2}{c}{$n=100$}\\
\hline
$H_{11}, X\sim N(0,\Sigma_1)$
& 0 &&0.0630 &&   0.0565 && 0.0470 &&   0.0500\\
& 0.2 && 0.0890 &&   0.1535 && 0.0730 &&   0.1263\\
& 0.4 && 0.2175 &&   0.4735 && 0.1623 &&   0.3857\\
& 0.6 && 0.4290 &&   0.8125 && 0.3207 &&   0.7227\\
& 0.8 && 0.6470 &&   0.9630 && 0.4910 &&   0.9207\\
& 1.0 && 0.8135 &&   0.9990 && 0.6347 &&   0.9803
\\
                          \hline
$X\sim N(0,\Sigma_2)$ & 0 && 0.0460 &&   0.0545 && 0.0480  &&  0.0563\\
& 0.2 && 0.0570  &&  0.1050 && 0.0767 &&   0.1173\\
& 0.4 && 0.1165  &&  0.3255 && 0.1647 &&   0.3667\\
& 0.6 && 0.2275  &&  0.6530 && 0.3243 &&   0.7203\\
& 0.8 && 0.4100  &&  0.8885 && 0.4953 &&   0.9293\\
& 1.0 && 0.5230  &&  0.9690 && 0.6787 &&   0.9887\\
                          \hline
$H_{12}, X\sim N(0,\Sigma_1)$ & 0 &&0.0535  &&  0.0610 && 0.0490  &&  0.0470\\
& 0.2 && 0.1275  &&   0.1845 && 0.0990 && 0.1687 \\
& 0.4 && 0.2950  &&   0.5695 && 0.2657 && 0.5510 \\
& 0.6 && 0.5715  &&   0.8895 && 0.5383 && 0.8980 \\
& 0.8 && 0.8100  &&   0.9945 && 0.7763 && 0.9890 \\
& 1.0 && 0.9410  &&   1.0000 && 0.9267 && 0.9993\\
                          \hline
$X\sim N(0,\Sigma_2)$ & 0 && 0.0395  &&  0.0380 && 0.0460  &&  0.0523\\
& 0.2 && 0.0700 &&  0.1160 && 0.0773 &&   0.1140\\
& 0.4 && 0.1545 &&  0.3690 && 0.1820 &&   0.3717 \\
& 0.6 && 0.3285 &&  0.6920 && 0.3660 &&   0.6953\\
& 0.8 && 0.5490 &&  0.9025 && 0.5723 &&   0.9130\\
& 1.0 && 0.7340 &&  0.9805 && 0.7587 &&   0.9857\\
                          \hline
\end{tabular}
\end{center}
}
\end{table}

\newpage

\begin{table}
\scriptsize{ \caption{Empirical sizes and powers of ${T}^{MAVE*}_n$ and $T^{DEE*}_n$ for $H_0$ vs.
$H_{11}$ and $H_{12}$, with $X\sim N(0,\Sigma_i)$, $i=1,2$ and $\epsilon\sim
N(0,1)$.}
 \label{tab2}
\begin{center}
\scriptsize
\begin{tabular}{cccccccccccccc}
\hline
 & $a$  & \multicolumn{4}{c}{$T^{MAVE*}_n$} & &  \multicolumn{4}{c}{$T^{DEE*}_n$}\\
 &&\multicolumn{2}{c}{$n=50$} &  \multicolumn{2}{c}{$n=100$} && \multicolumn{2}{c}{$n=50$} &\multicolumn{2}{c}{$n=100$}\\
\hline
$H_{11}, X\sim N(0,\Sigma_1)$
& 0 &&0.0697 &&   0.0580 && 0.0470 &&   0.0500\\
& 0.2 && 0.1160 &&   0.1890 && 0.0840  && 0.1425\\
& 0.4 && 0.2740 &&   0.5260 && 0.1635 &&   0.3900\\
& 0.6 && 0.4670 &&   0.8510 && 0.3255 &&   0.7235\\
& 0.8 && 0.6750 &&   0.9750 && 0.5115 &&   0.9185\\
& 1.0 && 0.8320 &&   0.9960 && 0.6160 &&   0.9780
\\
                          \hline
$X\sim N(0,\Sigma_2)$ & 0 && 0.0490 &&   0.0550 && 0.0500 &&   0.0475\\
& 0.2 && 0.0770 &&  0.1160 && 0.0680 &&  0.1285\\
& 0.4 && 0.1600 &&  0.3460 && 0.1515 &&   0.3480\\
& 0.6 && 0.2850 &&  0.7170 && 0.3080 &&   0.7155\\
& 0.8 && 0.4520 &&  0.8930 && 0.4835 &&  0.9255\\
& 1.0 && 0.5800 &&  0.9740 && 0.6660 &&   0.9820\\
                          \hline
$H_{12}, X\sim N(0,\Sigma_1)$ & 0 &&0.0770 &&   0.0740 && 0.0435 && 0.0555\\                          & 0.2 && 0.1520  &&  0.2070 && 0.1095 && 0.1680\\
& 0.4 && 0.3320 &&   0.5990 && 0.2545 && 0.5510\\
& 0.6 && 0.6170 &&   0.9200 && 0.5600 && 0.8975\\
& 0.8 && 0.8410 &&   0.9990 && 0.7690 && 0.9925\\
& 1.0 && 0.9430 &&   1.0000 && 0.9215 && 1.0000
\\
                          \hline
$X\sim N(0,\Sigma_2)$ & 0 && 0.0570  &&   0.058 && 0.0420 && 0.0540\\
& 0.2 && 0.0990 &&  0.1250 && 0.0705 && 0.1275\\
& 0.4 && 0.2070 &&  0.3560 && 0.1770 && 0.3495 \\
& 0.6 && 0.3710 &&  0.7070 && 0.3380 && 0.6870\\
& 0.8 && 0.5920 &&  0.9150 && 0.5390 && 0.9190\\
& 1.0 && 0.7600 &&  0.9810 && 0.7445 && 0.9805\\
                          \hline
\end{tabular}
\end{center}
}
\end{table}

\begin{table}
\scriptsize{ \caption{Empirical sizes and powers for $H_0$ vs.
$H_{13}$, with $X\sim N(0,\Sigma_i)$, $i=1,2$ and $\epsilon\sim
N(0,1)$.}
 \label{tab3}
\begin{center}
\scriptsize
\begin{tabular}{cccccccccccccc}
\hline
 & $a$  & \multicolumn{4}{c}{$\widetilde{T}^{MAVE}_n$} & &  \multicolumn{4}{c}{$T^{DEE}_n$}\\
 &&\multicolumn{2}{c}{$n=50$} &  \multicolumn{2}{c}{$n=100$} && \multicolumn{2}{c}{$n=50$} &\multicolumn{2}{c}{$n=100$}\\
\hline
$X\sim N(0,\Sigma_1), \epsilon\sim N(0,1)$ & 0 &&0.0515 &&   0.0625 && 0.0507 &&   0.0563\\
 & 0.2 && 0.1255 &&   0.2245 && 0.1067 && 0.2000\\
 & 0.4 && 0.3465 &&   0.7520 && 0.3330 && 0.7127\\
 & 0.6 && 0.6390 &&   0.9790 && 0.6170 && 0.9580\\
 & 0.8 && 0.8335 &&   0.9980 && 0.8023 && 0.9960\\
 & 1.0 && 0.9240 &&   1.0000 && 0.8897 && 0.9993\\
                          \hline
$X\sim N(0,\Sigma_2), \epsilon\sim N(0,1)$ & 0 &&0.0485 &&   0.0505 && 0.0477 &&   0.0480\\
& 0.2 && 0.4160  &&  0.8745 && 0.4163  &&  0.8523\\
& 0.4 && 0.8760  &&  0.9995 && 0.8933  &&  0.9993\\
& 0.6 && 0.9515  &&  1.0000 && 0.9680  &&  0.9997\\
& 0.8 && 0.9750  &&  1.0000 && 0.9860  &&  1.0000\\
& 1.0 && 0.9765  &&  1.0000 && 0.9907  &&  1.0000\\
                          \hline
\hline
 & $a$  & \multicolumn{4}{c}{$T^{MAVE*}_n$} & &  \multicolumn{4}{c}{$T^{DEE*}_n$}\\
 &&\multicolumn{2}{c}{$n=50$} &  \multicolumn{2}{c}{$n=100$} && \multicolumn{2}{c}{$n=50$} &\multicolumn{2}{c}{$n=100$}\\
\hline
$X\sim N(0,\Sigma_1), \epsilon\sim N(0,1)$ & 0 &&0.0767 &&   0.0640 && 0.0490 && 0.0490\\
 & 0.2 && 0.1580  &&  0.2570 && 0.1075 && 0.1835\\
 & 0.4 && 0.3860 &&   0.7190 && 0.3120 && 0.6800\\
 & 0.6 && 0.6170 &&   0.9560 && 0.5655 && 0.9410\\
 & 0.8 && 0.7940 &&   0.9920 && 0.7400 && 0.9885\\
 & 1.0 && 0.8660 &&   0.9930 && 0.8475 && 0.9930\\
                          \hline
$X\sim N(0,\Sigma_2), \epsilon\sim N(0,1)$ & 0 &&0.0580 &&   0.0435 && 0.0400 && 0.0475\\
& 0.2 && 0.4170  &&  0.8310 && 0.3925 && 0.7855\\
& 0.4 && 0.7950 &&   0.9880 && 0.7755 && 0.9870\\
& 0.6 && 0.8770 &&   0.9920 && 0.8985 && 0.9955\\
& 0.8 && 0.8870 &&   1.0000 && 0.9390 && 1.0000\\
& 1.0 && 0.8880 &&   1.0000 && 0.9445 && 1.0000\\
                          \hline
\end{tabular}
\end{center}
}
\end{table}

\clearpage

\begin{table}
\scriptsize{ \caption{Empirical sizes and powers in \emph{Study 3}, with $p=2$.
Here cases 1-4 respectively represent situations with
$X\sim N(0,\Sigma_1)$, $\epsilon\sim N(0,1)$; (case 1) or $DE(0,\sqrt{2}/2)$ (case 2) and
$X\sim N(0,\Sigma_2)$, $\epsilon\sim N(0,1)$ (case 3) or $DE(0,\sqrt{2}/2)$ (case 4) respectively.}}
\label{table4}
\begin{center}
\scriptsize
\begin{tabular}{ccccccccccccccccccc}
\hline
 & $a$  & \multicolumn{4}{c}{$T^{ZH}_n$} & \multicolumn{4}{c}{$T^{ZH*}_n$} &  \multicolumn{4}{c}{$T^{DEE}_n$} &  \multicolumn{4}{c}{$T^{DEE*}_n$}\\
 &&\multicolumn{2}{c}{$n=50$} &  \multicolumn{2}{c}{$n=100$} & \multicolumn{2}{c}{$n=50$} &  \multicolumn{2}{c}{$n=100$} & \multicolumn{2}{c}{$n=50$} &\multicolumn{2}{c}{$n=100$} & \multicolumn{2}{c}{$n=50$} &\multicolumn{2}{c}{$n=100$}\\
\hline
$\mathrm{Case} 1$ & 0 && 0.0470 &&  0.0375 && 0.0500  &&  0.0490    && 0.0483 &&   0.0453
  && 0.0463   &&  0.0527\\
 & 0.2 && 0.0820 &&   0.1390 && 0.0905 &&   0.1335  && 0.1523 &&   0.2653 && 0.1390 &&    0.2683 \\
 & 0.4 && 0.2430 &&   0.5065 && 0.2560  &&   0.4885   && 0.4127  &&  0.7610 && 0.4053 &&    0.7530  \\
 & 0.6 && 0.5175  &&  0.8635 &&  0.4245  &&  0.8320  &&0.6747   && 0.9650 &&  0.6573  &&   0.9557\\
 & 0.8 && 0.7335  &&  0.9825 && 0.6350   && 0.9465    &&  0.8453  &&  0.9943 && 0.8237  &&   0.9933  \\
 & 1.0 && 0.8875  &&  0.9990  && 0.7175  &&  0.9750   &&  0.9227 &&   1.0000 &&  0.9010 &&    1.0000  \\
                          \hline
$\mathrm{Case} 2$& 0 && 0.0410 &&   0.0375  && 0.0370 &&   0.0585  && 0.0497 &&    0.0487 &&  0.0453  &&  0.0530 \\
 & 0.2 && 0.0915 &&   0.1260  && 0.1005   &&   0.1485    && 0.1653  &&   0.2867 && 0.1540  &&   0.2867  \\
 & 0.4 && 0.2725 &&   0.5300    &&  0.2630  &&   0.5035    && 0.4293  &&   0.7540 && 0.4213  &&  0.7683   \\
 & 0.6 && 0.5460 &&   0.8635    &&  0.4840   &&   0.8335  && 0.6757  &&   0.9593 &&  0.6643  &&  0.9557  \\
 & 0.8 && 0.7565 &&   0.9790  &&  0.6115    &&   0.9520    && 0.8370 &&   0.9927 && 0.8287  &&   0.9947   \\
 & 1.0 && 0.8745 &&   0.9990  &&    0.7220    &&  0.9675   && 0.9183 &&    0.9978 && 0.9000 && 0.9965   \\                                             \hline
$\mathrm{Case} 3$ & 0 && 0.0370  &&  0.0405  && 0.0435 &&   0.0590   && 0.0473  &&   0.0527 &&  0.0597  &&  0.0513   \\
 & 0.2  && 0.0830  &&  0.1130   &&   0.0945  &&  0.1760   && 0.1237 &&    0.2107 && 0.1087  &&  0.2130   \\
 & 0.4  && 0.2630  &&  0.5340 &&  0.2705   && 0.5220     && 0.3667  &&   0.6417 &&0.3320  &&  0.6410 \\
 & 0.6  && 0.5295 &&   0.8955  &&  0.4910  &&  0.8505   && 0.6220  &&   0.9363 && 0.5913 &&   0.9250 \\
 & 0.8  && 0.7915 &&   0.9855 && 0.6385  &&  0.9580    && 0.8170  &&   0.9923 && 0.7797  &&  0.9870  \\
 & 1.0  && 0.9115 &&   0.9995 &&  0.7300   && 0.9775   && 0.9183  &&   0.9990 &&  0.8757  &&  0.9987  \\
                          \hline
 $\mathrm{Case} 4$& 0 && 0.0370  &&  0.0410  && 0.0455  &&  0.0420   && 0.0463 &&   0.0513 && 0.0565  &&  0.0565  \\
 & 0.2 && 0.0890 &&   0.1495   &&  0.0995  &&  0.1540   && 0.1367  &&  0.2140 &&  0.1305  &&  0.2105 \\
 & 0.4 && 0.2960 &&   0.5490  &&   0.2680 &&   0.5285   && 0.3677 &&   0.6623 &&  0.3690  &&  0.6695\\
 & 0.6 && 0.5750 &&   0.8955  &&  0.5060   &&   0.8500   && 0.6450  &&  0.9333 && 0.5960  &&  0.9175 \\
 & 0.8 && 0.7885 &&   0.9895 && 0.6600  &&  0.9510    && 0.8147 &&   0.9897 && 0.7725 &&   0.9825\\
 & 1.0 &&  0.9005 &&   0.9980 &&  0.7380  &&  0.9790   && 0.9110 &&   0.9997 &&   0.8855 &&   0.9980\\ \hline
\end{tabular}
\end{center}
\end{table}

\begin{table}
\scriptsize{ \caption{Empirical sizes and powers in \emph{Study 3}, with $p=8$. Here cases 1-4 represent the situations with
$X\sim N(0,\Sigma_1)$, $\epsilon\sim N(0,1)$ (case 1) or $DE(0,\sqrt{2}/2)$ (case 2) and
$X\sim N(0,\Sigma_2)$, $\epsilon\sim N(0,1)$ (case 3) or $DE(0,\sqrt{2}/2)$ (case 4) respectively.}
 \label{tab5}
\begin{center}
\scriptsize
\begin{tabular}{ccccccccccccccccccc}
\hline
 & $a$  & \multicolumn{4}{c}{$T^{ZH}_n$} & \multicolumn{4}{c}{$T^{ZH*}_n$} &  \multicolumn{4}{c}{$T^{DEE}_n$} &  \multicolumn{4}{c}{$T^{DEE*}_n$}\\
 &&\multicolumn{2}{c}{$n=50$} &  \multicolumn{2}{c}{$n=100$} & \multicolumn{2}{c}{$n=50$} &  \multicolumn{2}{c}{$n=100$} & \multicolumn{2}{c}{$n=50$} &\multicolumn{2}{c}{$n=100$} & \multicolumn{2}{c}{$n=50$} &\multicolumn{2}{c}{$n=100$}\\
\hline
$\mathrm{Case} 1$ & 0 && 0.0182 && 0.0297 && 0.0450  &&  0.0415   && 0.0605 &&0.0460 && 0.0465  &&  0.0495\\
 & 0.2 && 0.0280 &&   0.0475 && 0.0500  &&  0.0705  && 0.1400 &&   0.2645 && 0.1345 &&   0.2610\\
 & 0.4 && 0.0442 &&   0.0795 && 0.0785  &&  0.0930  && 0.3485 &&   0.6990 && 0.3555 &&   0.7145\\
 & 0.6 && 0.0742  &&  0.1573 && 0.1035  &&  0.1895  && 0.5905 &&   0.9420 && 0.5555 &&   0.9280\\
 & 0.8 && 0.1022 &&   0.2627 && 0.1330  &&  0.2770  && 0.7510 &&   0.9855 && 0.7275 &&   0.9890\\
 & 1.0 && 0.1422 &&   0.3715  && 0.1630  &&  0.3500  && 0.8475 &&   0.9960 && 0.8170 &&   0.9935\\
                          \hline
$\mathrm{Case} 2$& 0 && 0.0175 &&   0.0265  &&   0.0480  &&  0.0430   && 0.0543    && 0.0517 && 0.0400  &&  0.0540\\
 & 0.2 && 0.0283 &&   0.0508  &&  0.0655 &&   0.0590  && 0.1463 &&   0.2843 && 0.1395 &&   0.2760\\
 & 0.4 && 0.0522  &&  0.0953     &&   0.0865 &&   0.1240  && 0.3740 &&   0.7240 && 0.3610 &&   0.7365\\
 & 0.6 && 0.0885  &&  0.1955    &&    0.1295 &&   0.2070  && 0.6073 &&   0.9323 && 0.5990 &&   0.9260\\
 & 0.8 && 0.1323 &&   0.2953   &&    0.1560 &&   0.2925  && 0.7470 &&   0.9873 && 0.7355 &&   0.9860\\
 & 1.0 && 0.1675 &&   0.4070  &&     0.1880 &&   0.3955  && 0.8510 &&   0.9980 && 0.8170 &&   0.9935\\
                          \hline
$\mathrm{Case} 3$ & 0 && 0.0213 &&   0.0280  &&  0.0480  &&  0.0485  && 0.0463 &&   0.0483 && 0.0450 &&   0.0590\\
 & 0.2 && 0.0600 &&   0.1335   &&  0.0765 &&   0.1660  && 0.3237  &&0.6443 && 0.2905 &&   0.6595\\
 & 0.4 && 0.1935 &&  0.4572 &&  0.2245 &&   0.4265  && 0.6970  &&0.9797 && 0.6780 &&   0.9780\\
 & 0.6 && 0.3445 &&   0.7280  && 0.3220 &&   0.6570  && 0.8773  &&0.9993 && 0.8340 &&   0.9980\\
 & 0.8 && 0.4758 &&   0.8588 && 0.4290 &&   0.7535  && 0.9237  &&0.9993 && 0.8985 &&   0.9990\\
 & 1.0 && 0.5480 &&   0.9205 && 0.4750 &&   0.8020  && 0.9527  &&1.0000 && 0.9335 &&   0.9990\\
                          \hline
 $\mathrm{Case} 4$& 0 && 0.0190  &&  0.0275  &&  0.0460 &&   0.0540  && 0.0507    && 0.0533 && 0.0475    && 0.0500\\
 & 0.2 && 0.0742 &&   0.1465   &&     0.1095 &&   0.1915  && 0.3280 &&   0.6583 && 0.3140 &&   0.6600\\
 & 0.4 && 0.2350  &&  0.4950  &&   0.2370 &&   0.4420  && 0.6977 &&   0.9783 && 0.6975 &&   0.9760\\
 & 0.6 && 0.3757 &&   0.7445  &&  0.3440 &&   0.6495  && 0.8660 &&   0.9993 && 0.8355 &&   0.9985\\
 & 0.8 && 0.4765 &&   0.8585 && 0.4285 &&   0.7425  && 0.9250 &&   0.9997 && 0.9110 &&  1.0000\\
 & 1.0 && 0.5537  && 0.9147 && 0.4865 &&   0.8095  && 0.9550 &&   1.0000 && 0.9355 &&   1.0000\\ \hline
\end{tabular}
\end{center}
}
\end{table}

\clearpage
%

\end{document}